\documentclass[aps,prb,showkeys]{revtex4}
\usepackage[utf8]{inputenc}
\usepackage[T1]{fontenc}

\usepackage{booktabs}
\usepackage{amssymb}
\usepackage{amsmath}
\usepackage{graphicx}
\usepackage{float}
\usepackage{siunitx}
\usepackage{adjustbox}
\usepackage{amsmath}
\usepackage{empheq}
\usepackage[most]{tcolorbox}
\usepackage{makecell}
\begin{document}

\title{Entropy measures as indicators of connectivity paths in the human brain}

\author{Ania \surname{Mesa-Rodríguez}}
\email{ania@pks.mpg.de}
\affiliation{Max Planck Institute for the Physics of Complex Systems, Nonlinear Dynamics and Time Series Analysis, Dresden, 01187, Germany}
\affiliation{University of Havana, Physics and Mathematics Faculties, La Habana, 10400, Cuba}

\author{Ernesto \surname{Estevez-Rams}}
\affiliation{University of Havana, Physics and Mathematics Faculties, La Habana, 10400, Cuba}

\author{Holger \surname{Kantz}}
\affiliation{Max Planck Institute for the Physics of Complex Systems, Nonlinear Dynamics and Time Series Analysis, Dresden, 01187, Germany}

\begin{abstract}
We develop an analytical framework that benefits from a set of information theory variables to study brain activity under different stimuli. fMRI signals from different brain regions are treated as time series, and information production as well as pattern redundancy are measured using entropy density, effective measure complexity, and informational distance estimated by Lempel-Ziv complexity. This enables the detection of both linear and non-linear dynamics without relying on pre-established parameters, models, or prior assumptions about the data. The framework is applied to task-based fMRI data from the Human Connectome Project under motor, working memory, emotion recognition, and language tasks, as well as resting state. The complexity entropy map identifies regional engagement consistent with established task-relevant areas, providing a model-free indicator of activation. The Lempel-Ziv distance between region pairs is then used to construct distance matrices, dendrograms, and connectivity graphs that recover hierarchical and modular functional structure: active regions cluster by functional specialization, while non-active regions form a densely interconnected backbone shared across tasks. Since the analytical framework does not depend on prior knowledge, it is well suited for exploratory research and facilitates the discovery of previously unreported connections or patterns in the brain activity. The capacity to identify non-linear dynamics is especially important for studying brain connectivity, as the brain exhibits significant non-linear interactions across multiple functional levels.
\end{abstract}


\date{\today}

\flushbottom

\maketitle

\section*{Introduction}

The topological and functional complexity of the brain is a crucial aspect of its operation; its ability to encode information is not limited to a single scale, ranging from macroscopic to microscopic levels \cite{swansonCajalConnectome2016}. Although a neuron is considered the primary computational unit essential for brain function at the microscale, cortical columns composed of anatomically or functionally distinct neuronal populations represent the minimal intelligent units capable of performing logical operations at the mesoscopic scale, which is closely related to noninvasive recordings \cite{spieglerDynamicsBiologicallyInformed2012}. At the macroscopic level, techniques such as fMRI and structural MRI provide insights into the organization of the whole brain, large-scale networks, and functional specialization in areas of the brain. Recent studies have attempted to bridge the multiple scales of organization into an integrative theory, finding consistent organizational principles of connectivity at both the macroscale and the neuronal levels \cite{scholtensCommonMicroscaleMacroscale2022, vandenheuvelBridgingCytoarchitectonicsConnectomics2015}.

The human connectome represents the intricate network of structural and functional connections between different regions and areas of the brain, providing crucial insights into how these regions are interconnected and communicate~\cite{sporns2005human, hagmann2005diffusion}. This understanding is fundamental to unraveling brain function, including processes such as cognition, perception, emotion, and memory. Mapping the connectome can help identify abnormal connectivity patterns associated with conditions such as Alzheimer's disease \cite{yuHumanConnectomeAlzheimer2021,yanRichClubDisturbances2018}, Parkinson \cite{bellucciReviewParkinsonDisease2016}, schizophrenia, ataxia, autism, and traumatic brain injuries \cite{foxMappingSymptomsBrain2018}. This knowledge is vital for diagnosis, treatment, and potential interventions. In fMRI measurements, when performing cognitive tasks that require attention, a change in response is observed in different regions, some signaling increases in activity, some decrease in activity, and some remain more or less indifferent \cite{cab00,gusnard2001searching,simpson2001emotion,corbetta2002control}.

When analyzing fMRI data different approaches have been used to infer direct interactions between brain regions through various methods, including causal models \cite{valdes-sosaEffectiveConnectivityInfluence2011,bielczykThresholdingFunctionalConnectomes2018} which often requires strong model assumptions and are not well-suited for exploratory analyses; Granger causality and the WAGS framework \cite{barnettGrangerCausalityState2015, valdes-sosaEffectiveConnectivityInfluence2011} which assume linearity, stationarity and cannot detect instantaneous or non-linear causal effects; the Directed Transfer Function (DTF) \cite{kaminskiNewMethodDescription1991} and the closely related Partial Directed Coherence (PDC) \cite{baccalaPartialDirectedCoherence2001}, two frequency-domain measures built on the same multivariate autoregressive formulation and belonging to the Granger-causal framework, which therefore share its assumptions of linearity and stationarity and are additionally sensitive to model order and parameter estimation; correlations and covariance, which only capture linear associations and are confounded by indirect connections \cite{schiefer2018correlation}; graph theory \cite{bullmoreComplexBrainNetworks2009} which is highly sensitive to threshold selection; and community detection \cite{akikiDeterminingHierarchicalArchitecture2019} which depends on the chosen algorithm and may miss overlapping or hierarchical communities, among others.

What is still lacking is a data analysis approach that allows, without relying on parameters, models, or correct prior knowledge of the network, to characterize  cognitive states and to recover functional connections between brain regions, detecting both linear and non-linear dynamics. This article reports on such data pipeline analysis. That such an approach is relevant is witnessed by the already explained shortcomings of previous efforts. Specially relevant is not to assume linear responses or correlation between regions, or even that non-linearity can be accommodated as a perturbation of an otherwise linear response. Our analysis will show the robustness of the quantitative method, applicable to a wide range of experimental conditions and signals. Although we tested the procedure on fMRI data extracted from the Human Connectome Project \cite{vanessenHumanConnectomeProject2012}, it is straightforward to generalize it to other types of signal, such as the electroencephalogram. Task-based fMRI time series from subjects performing motor, emotion, working memory, and language tasks, as well as resting-state, were used.

The procedure reported is based on information theory. Shannon entropy has been used before to analyze brain activity \cite{tianBridgingInformationDynamics2021, wangBrainEntropyMapping2014}, yet previous use of such magnitudes has relied solely on the use of entropy density. In our approach, a broader range of information theory tools, including entropy density, effective measure complexity, and informational distance, are used to assess creativity and the emergence of patterns. Their joint use in the $(h, E)$ map provides an indicator of regional engagement under each task that requires no generative model, no design matrix, and no tuned parameters, the only fixed choices being the binarization threshold and the neighborhood size of the activation criterion, whose influence is shown to be minor (Supplementary Section~3). An informational distance between pairs of regions is introduced to recover functional connectivity. The distance quantifies how closely two signals match in terms of one reproducing the patterns of the other, independent of time lag, and is used here to build distance matrices, hierarchical dendrograms, and connectivity graphs across the 360 cortical regions of the Glasser parcellation. The ability to detect non-linear dynamics is particularly relevant for analyzing brain connectivity, as the brain exhibits strong non-linear interactions across different levels of function \cite{gautamaIndicationsNonlinearStructures2003}.

The entropy density, $h$, which is also known as the entropy rate or Kolmogorov-Sinai entropy depending on the field, is an intensive measure of unpredictability in a signal, measured by the signal's ability to create new patterns \cite{coverELEMENTSINFORMATIONTHEORY}; while structure in a signal can be assessed by the effective measure complexity, $E$, also known as excess entropy, related to the persistence of patterns or memory of a system \cite{grassberger86,crutchfieldRegularitiesUnseenRandomness2003}. An interesting previous use of similar ideas has been reported in the study of written language \cite{estevez-ramsComplexityentropyAnalysisDifferent2019}. Language exhibits a complex balance between (grammatical and logical) rules and the production of information. Entropy density was reported to be tied to the creativity of a text, the capacity of a text to create information as we move along the character sequence. Effective measure complexity, on the other hand, captures the redundancy of the text, a needed characteristic of an information source if it intends to convey useful information. The relevance of $E$ is that it grasps this reliance on memory at all scales in a single parameter. The hallmark of complexity lies between total unpredictability, or maximum entropy density value, on one side, and complete predictability or zero entropy density, on the other. Extending such ideas to the signals coming from the brain regions, one hopes that their complex behavior can be captured, at least partially, by the tuple $(h, E)$, and that is the hypothesis tested in this work. Furthermore, an additional magnitude, informational distance, $d$, is also introduced as relevant when studying functional connections between regions. Informational distance measures how distant one signal is from another, in terms of one reproducing the patterns of the other, independent of time lag, while possibly generating its own independent behavior.

\section*{Methods}\label{sec2}

\subsection*{Data acquisition.}\label{subsec2.1}
 Task-based fMRI time series taken from the Human Connectome Project data server \cite{vanessenHumanConnectomeProject2012} were used for five types of exercises: resting state, motor, working memory, emotion, and language; details can be found in Wu et. al\cite{wu-minn1200SubjectsData2017}.

During resting-state fMRI (rfMRI) scans, participants kept their eyes open, maintaining a relaxed fixation on a bright cross-hair projected onto a dark background in a dark room.

For the motor task, participants received visual cues instructing them to either tap their left or right fingers, squeeze their left or right toes, or move their tongue. Each movement block lasted 12 seconds (consisting of 10 movements) and was preceded by a 3-second visual cue. Across two runs, there were 13 blocks, including 2 blocks of tongue movements, 4 blocks of hand movements (2 right and 2 left), and 4 blocks of foot movements (2 right and 2 left). Additionally, each run included 3 fixation blocks of 15 seconds.

In the working memory exercise, participants were shown blocks of trials featuring images of places, tools, faces, and body parts (non-mutilated and non-nude). During each run, these four types of stimuli were presented in separate blocks; half of the blocks involved a 2-back working memory task, and the other half used a 0-back task as a comparison for working memory performance. At the start of each block, a 2.5-second cue indicated the task type (and the target for the 0-back task). Each of the two runs included 8 task blocks (comprising 10 trials of 2.5 seconds each, lasting 25 seconds per block) and 4 fixation blocks (lasting 15 seconds). For each trial, the stimulus was displayed for 2 seconds, followed by a 500 ms inter-task interval (ITI). 

In the emotion task, participants are shown blocks of trials where they are asked to either identify which of two faces at the bottom of the screen matches the face at the top, or which of two shapes at the bottom matches the shape at the top. The faces display either an angry or fearful expression. The trials are grouped into blocks of 6 trials focused on either faces or shapes. Each stimulus is shown for 2000 ms, followed by a 1000 ms inter-trial interval. Before each block, a 3000 ms cue indicates whether the task is "shape" or "face," making the total block duration 21 seconds, including the cue. Each of the two runs contains 3 face blocks and 3 shape blocks, with 8 seconds of fixation at the end of each run.

Finally, the language task involves two runs, each alternating between 4 blocks of a story task and 4 blocks of a math task. The block lengths vary, averaging around 30 seconds, but the math blocks are designed to match the length of the story blocks, with some additional math trials at the end of the task if needed to complete the 3.8-minute duration. In the story blocks, participants listen to short auditory stories (5-9 sentences) followed by a two-choice question that tests their understanding of the story's theme. The math task also presents auditory trials in which participants are asked to solve addition or subtraction problems. Each trial presents a series of arithmetic operations followed by two choices for selecting the correct answer. The math task is adaptive, adjusting the difficulty level to maintain consistent challenge across participants.

The data was preprocessed through the Human Connectome Project standard preprocessing pipeline, which ensures that the fMRI data is cleaned and standardized, making it suitable for analysis. This includes essential steps such as motion correction to reduce distortions caused by participant movement during scanning, removal of noise and artifacts, alignment to a standard brain template to ensure that brain regions are consistently mapped across subjects, spatial smoothing, and temporal filtering, among others (Details can be found in Wu et. al.\cite{wu-minn1200SubjectsData2017}).

The brain was divided into 360 regions of interest (ROIs) using the Glasser atlas \cite{glasserMinimalPreprocessingPipelines2013}. The Glasser parcellation covers the cerebral cortex only; subcortical structures, including the amygdala, hippocampus, thalamus and basal ganglia, are not represented. Task components whose principal substrate is subcortical are therefore outside the reach of this analysis, a limitation most relevant to the emotion task (amygdala, hippocampus) and discussed where it bears on the results. ROIs 1-180 correspond to the left hemisphere; ROIs 181-360 are their homologous counterparts in the right hemisphere.

\subsection*{Data preprocessing.}\label{subsec2.2}
\begin{figure}[ht!]
        \begin{center}
               \includegraphics[scale=0.7]{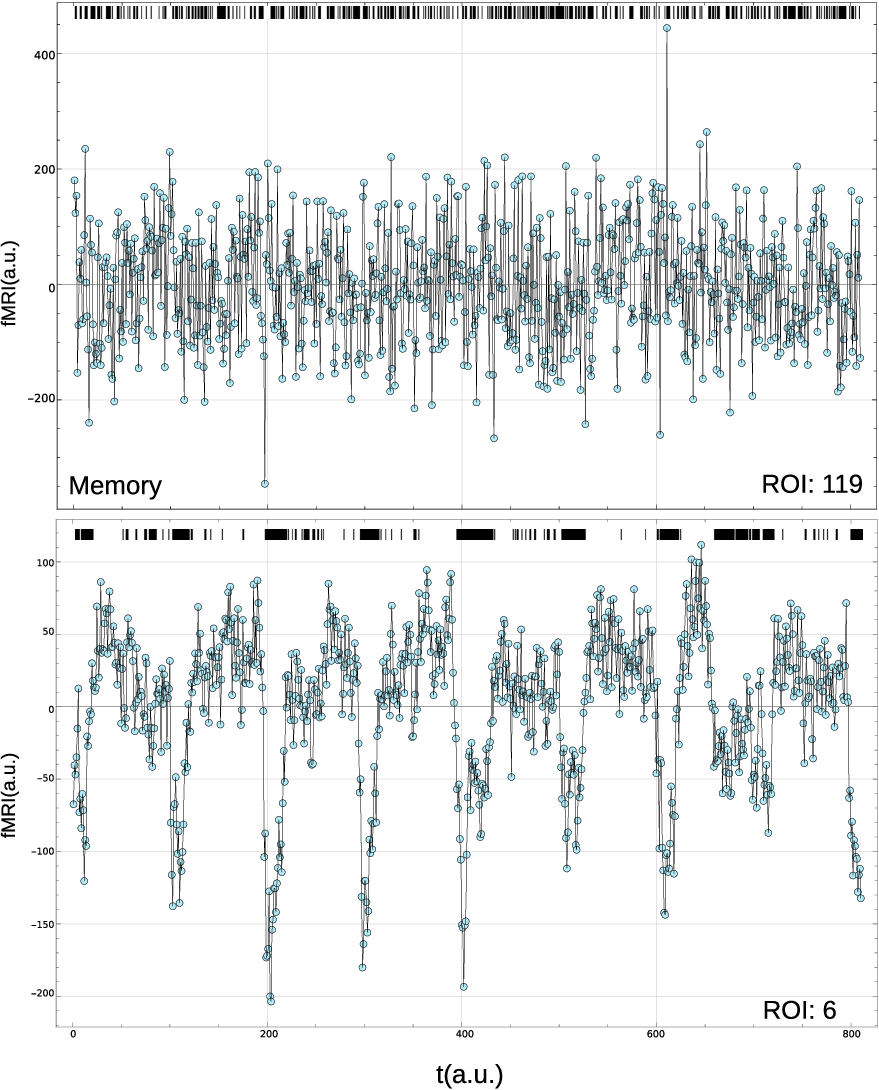}
        \end{center}
                \caption{\small{\textbf{fMRI signal.} Blood oxygenation level-dependent (BOLD) signal from the fMRI measurement of (upper) inactive region, and (lower) active region of the brain. Binarization (black-and-white bar over the signal plot) is performed using the mean value as a threshold. The binary sequence shows, as a time series, if the signal is below (0) or above (1)  the threshold value, for a given instant of time, which can be related, referred to as the baseline, to the higher or lower oxygen content of the region, in turn indirectly related to the region's activity.
                }}
            \label{fig:timeseries}
\end{figure}

 fMRI measures the Blood Oxygenation Level-Dependent (BOLD) signal, which reflects changes in blood oxygen levels. The principle behind it is that when a brain region becomes active, it consumes more oxygen and glucose, leading to increased blood flow to that area. This change is detected by MRI due to the magnetic properties of deoxyhemoglobin. (Figure \ref{fig:timeseries}).

The BOLD signal is an indirect, vascular proxy of neural activity: neural events are convolved with the hemodynamic response function (HRF), which acts as a low-pass filter with a time to peak that varies by a few seconds across cortical regions and subjects \cite{handwerkerVariationBOLDHemodynamic2004, aguirreVariabilityHumanBOLD1998}. This delay, and its inter-regional variability, is a well-known complication for connectivity methods that rely on temporal precedence, such as Granger causality, since differences in HRF timing can reverse the apparent order of activation between two regions \cite{rangaprakashHemodynamicResponseFunction2018, davidIdentifyingNeuralDrivers2008}. The framework used here is largely insensitive to this issue. The entropic measures $h_{LZ}$ and $E_{LZ}$ depend on the pattern statistics of the binarized signal over the full recording window rather than on the timing of individual events, and the LZ-distance is by construction time-lag independent, capturing shared pattern content regardless of when those patterns occur in each sequence. The slow-envelope structure of task-related activity that shapes these statistics is preserved by the hemodynamic filter, and no HRF deconvolution is therefore applied; this also preserves the model-free character of the approach. The consistency of the activation maps with established task-relevant areas, reported in the following sections, supports this argument empirically.

Before the entropic measures are applied, the data needs to be discretized. For each time series, the mean value is computed and used as a threshold for binarization. Binarization is a drastic procedure resulting in the loss of information from the original signal. What is hoped for is that the binary sequence still carries enough information to grasp relevant features of the dynamics of the system. The choice of the mean as the threshold, rather than the median or a quantile-based criterion, preserves a feature of the signal that varies genuinely across regions: the relative time spent above and below baseline. The median, by construction, forces the binarized sequence to contain equal numbers of 0s and 1s regardless of the underlying dynamics, fixing the marginal symbol distribution and removing this degree of freedom. We also restrict the analysis to binary discretization rather than ordinal (Bandt–Pompe) encoding because the Lempel-Ziv estimator converges to the entropy density at a rate that depends on the alphabet size, and the fMRI runs analyzed here are short relative to the alphabet size that ordinal encoding would require. To verify that the binarization does not destroy the dynamical information relevant to the analysis, we compared $h_{LZ}$ to the spectral entropy of each region computed directly on the continuous BOLD signal; the two measures agree across all tasks (Supplementary Figure~2), confirming that the pattern statistics captured by $h_{LZ}$ reflect a genuine property of the underlying signal rather than an artifact of thresholding. We further repeated the full analysis using the median as the binarization threshold; the entropic maps, the active/non-active classification, and the over/under-performance pattern are preserved (Supplementary material Section~3), confirming that the results do not depend on the specific choice of central threshold.

The binary sequence can be understood as follows. Taking the mean value as a baseline reference, the signal space has been segmented into two regions, above and below the baseline. A change of values $1\leftrightarrows 0$ means crossing from one region to the other. Patterns in the binary sequence, therefore, are to be understood in terms of this crossing. Any variation of the signal within one of the two regions is lost in the discretization process (Figure \ref{fig:timeseries}).

\subsection*{Entropic measures.}\label{subsec2.3}
Three entropic-based magnitudes will be used in the study: entropy density,  effective measure complexity, and information distance.

Entropy density has been defined in several equivalent ways; the entropy density is related to the rate of change of the Shannon block entropy with a time series length \cite{coverELEMENTSINFORMATIONTHEORY}:
\begin{equation}
h = \lim_{L \rightarrow \infty} \frac{H(L)}{L},
\label{density}
\end{equation}
where $H(L) = -\sum p(s^L) \log p(s^L)$ is the Shannon block entropy over the probabilities $p(s^L)$ of a given subsequence $s^L$ of length $L$ in the time series. The sum is taken over all possible sequences of length $L$  \cite{coverELEMENTSINFORMATIONTHEORY}. The entropy density measures the amount of randomness or unpredictability when observing an infinitely long time series. We estimate the entropy density through the Lempel-Ziv factorization due to the difficulty of estimating the entropy density for finite-length sequences, this estimator has been proven to be robust even for short sequences \cite{lesne09}. The estimated entropy density will be denoted by $h_{LZ}$.

The definition of the Lempel-Ziv Complexity \cite{lempelComplexityFiniteSequences1976} is based on the exhaustive factorization of a sequence, adding a new factor whenever a new pattern appears.

A factorization of a sequence $S$ of length $N$ is defined as its partition into disjoint blocks:
$$F(S) = S(1,l_1), S(l_1+1,l_2), S(l_2+1,l_3), ..., S(l_{m-1}+1,N)$$
The factorization $F(S)$ is called the Lempel-Ziv factorization or exhaustive history if each factor $S(l_{k-1} +1, l_k)$ is not a substring of the string $S(1, l_k -1)$, while $S(l_{k-1} + 1, l_k -1)$ is a substring of $S(1, l_k -2)$ (except, maybe, for the last factor).

The exhaustive history is unique for each sequence \cite{lempelComplexityFiniteSequences1976}, allowing us to define the Lempel-Ziv complexity of the sequence $S$, denoted as $C_{LZ}(S)$, as the number of factors in its exhaustive history.

For example, given the sequence $S = 11011101000011$, its exhaustive history is $F(S) = 1.10.111.010.0001.1$, where each factor is delimited by a dot. This factorization contains 6 elements, so $C_{LZ}(S) = 6$.

It can be demonstrated \cite{zivCodingTheoremsIndividual1978} that if $S$ is the output of an ergodic process with finite memory, then the following holds:
\begin{equation}
\label{clz}
\overline{\lim_{N \rightarrow \infty}} \frac{C_{LZ} (S)}{N / \log N} = h(S),
\end{equation}
where $h(S)$ is the entropy density. Equation (\ref{clz}) is valid in the infinite limit; however, in practice, it is used as a robust estimate of the entropy density\cite{lesne09}.

Even if some of the restrictions used to arrive at the above expression (\ref{clz}) are lifted, $h_{LZ}$ still measures the pattern production of a finite sequence and, in that regard, can be used.

The effective measure complexity is the mutual information between the two halves of a (infinite) time series \cite{grassberger86}. In this sense, it conveys the amount of information that one half has on the other half and therefore, measures correlations (memory) at all length scales without the need to assume linear behavior. Estimation of effective measure complexity is in general a difficult task \cite{crutchfieldRegularitiesUnseenRandomness2003}, if it is assumed that the time series has finite memory and can be taken as stochastic, then a used related magnitude, which will be called LZ-effective complexity, is based on the random shuffling of the data \cite{melchert15,estevez19}. This is the estimate used in this report and will be denoted by $E_{LZ}$.

For estimating the LZ-effective complexity, a random shuffle procedure\cite{melchert15} is used that follows from the expression:
\begin{equation}
 E_{LZ}=\sum\limits_{M=1}^{M_{max}}[h_{LZ}(S_{(M)})-h_{LZ}(S)].\label{eq:elz}
\end{equation}
$S_{(M)}$ is obtained by partitioning the string $S$ in non-overlapping blocks of length $M$ and performing a random shuffling of the blocks. This destroys all correlations between symbols for lengths larger than $M$ while keeping the symbol frequency. $M_{max}$ is chosen to avoid fluctuations. In spite that $E_{LZ}$ is not strictly equivalent to the effective complexity measure, it behaves similarly in a number of cases, including stochastic finite memory processes.

The information distance \cite{li04} between two sequences $S$ and $Q$ is defined as follows: 
\begin{equation}
d(S,Q)= \frac{K(SQ)-\min \{ K(S),K(Q) \}}{\max \{ K(S), K(Q)\}}\label{eq:dd}
\end{equation}
where $K(S)$ represents the Kolmogorov (algorithmic) complexity \cite{kolmogorov65} of the time series $S$. $K(S)$ is the length of the shortest program that generates the sequence $S$ \cite{kolmogorov65}. $SQ$ denotes the concatenation of both sequences. $d$ is normalized between $0$ and $1$, the latter taken as the largest attainable distance between the compared time series. The information distance does not measure how the two sequences differ point by point, it is not a Hamming type distance; instead, it quantifies their informational similarity,  determining the degree of correlation between them from an algorithmic perspective: it reflects the shortest program length required to transform one sequence into the other, capturing how much new information is needed to describe one sequence in terms of the other.

A practical alternative is also necessary for this distance metric; the usual procedure has been to use compression algorithms to estimate the Kolmogorov complexities involved in the use of equation (\ref{eq:dd}). Instead, in this research, an estimate of Kolmogorov randomness will be made with Lempel-Ziv factorization as described in \cite{estevez15} and will be called LZ-distance and denoted by $d_{LZ}$. $d_{LZ}$ is a measure of the common patterns between two sequences without regard to the position where those patterns occur in each string.

Finally, information distance $d(s,p)$ will be estimated also via Lempel-Ziv by
\begin{equation}
 d_{LZ}(S,Q)=\frac{C_{LZ}(SQ)-min\{C_{LZ}(S), C_{LZ}(Q)\}}{max\{C_{LZ}(S), C_{LZ}(Q)\}}.\label{eq:dlz}
\end{equation}
which will have the same interpretation than $d(S,Q)$ as much as the normalized equation~\ref{clz} estimates the entropy density.

\subsection*{Analysis pipeline}\label{subsec2.4}
\begin{figure*}[ht!]
        \begin{center}
               \includegraphics[width=0.9\textwidth]{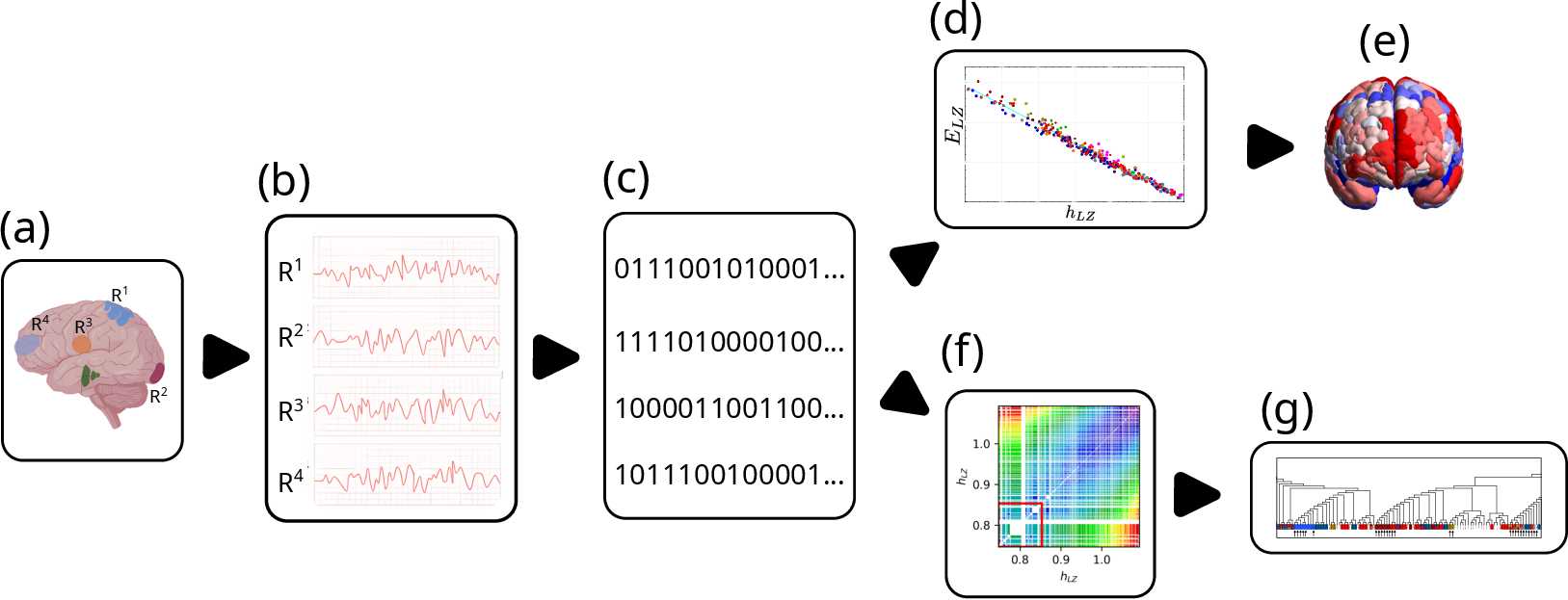}
        \end{center}
                \caption{\small{ \textbf{Analysis pipeline.} (a) The brain was divided into Regions of Interest (ROIs), and (b) the time series from the fMRI measurements of each region were (c) discretized and further analyzed using entropic measures, resulting in (d) the complexity-entropy maps of the task from where the (e) activated regions of the brain are determined. (f) The distance matrix between each pair of regions is computed, and the resulting (g) dendrogram of the region's distances is drawn.
                }}
            \label{fig:diagram}
\end{figure*}

The analysis followed in our study is depicted in Figure \ref{fig:diagram}: Under a given task, the fMRI signal from different regions of the brain according to Glasser parcellation is binarized; the entropic magnitudes from each resulting discrete sequence are estimated and the $E_{LZ}$ vs $h_{LZ}$, complexity-entropy, map is built, the activated regions of the brain are determined from the map. For every two regions of the brain, the informational distance $d_{LZ}$ is computed and a distance matrix is built; the distance map allows the building of the distance dendrogram.

\section*{Results and Discussion}\label{sec3}

The entropic map represents each brain region as a point in the $(h_{LZ}, E_{LZ})$ plane, where $h_{LZ}$ is the Lempel-Ziv entropy density and $E_{LZ}$ is the LZ-effective measure complexity, both computed from the binarized BOLD time series. This representation provides a compact characterization of each region's dynamics: $h_{LZ}$ quantifies the unpredictability of the signal, while $E_{LZ}$ measures the degree of pattern formation. Lower $h_{LZ}$ and higher $E_{LZ}$ indicate more structured, less random dynamics, the signature of task-driven activity that suppresses background noise~\cite{itoTaskevokedActivityQuenches2020}. We use active and activation in a specific sense: a region is active when its whole run BOLD dynamics are more structured and less random (lower $h_{LZ}$, higher $E_{LZ}$). This is distinct from the conventional fMRI sense of activation as response amplitude against a task-timed model; the two quantities are compared explicitly in the validation section.

\begin{figure*}[ht!]
        \centering
        		\includegraphics[width=\textwidth]{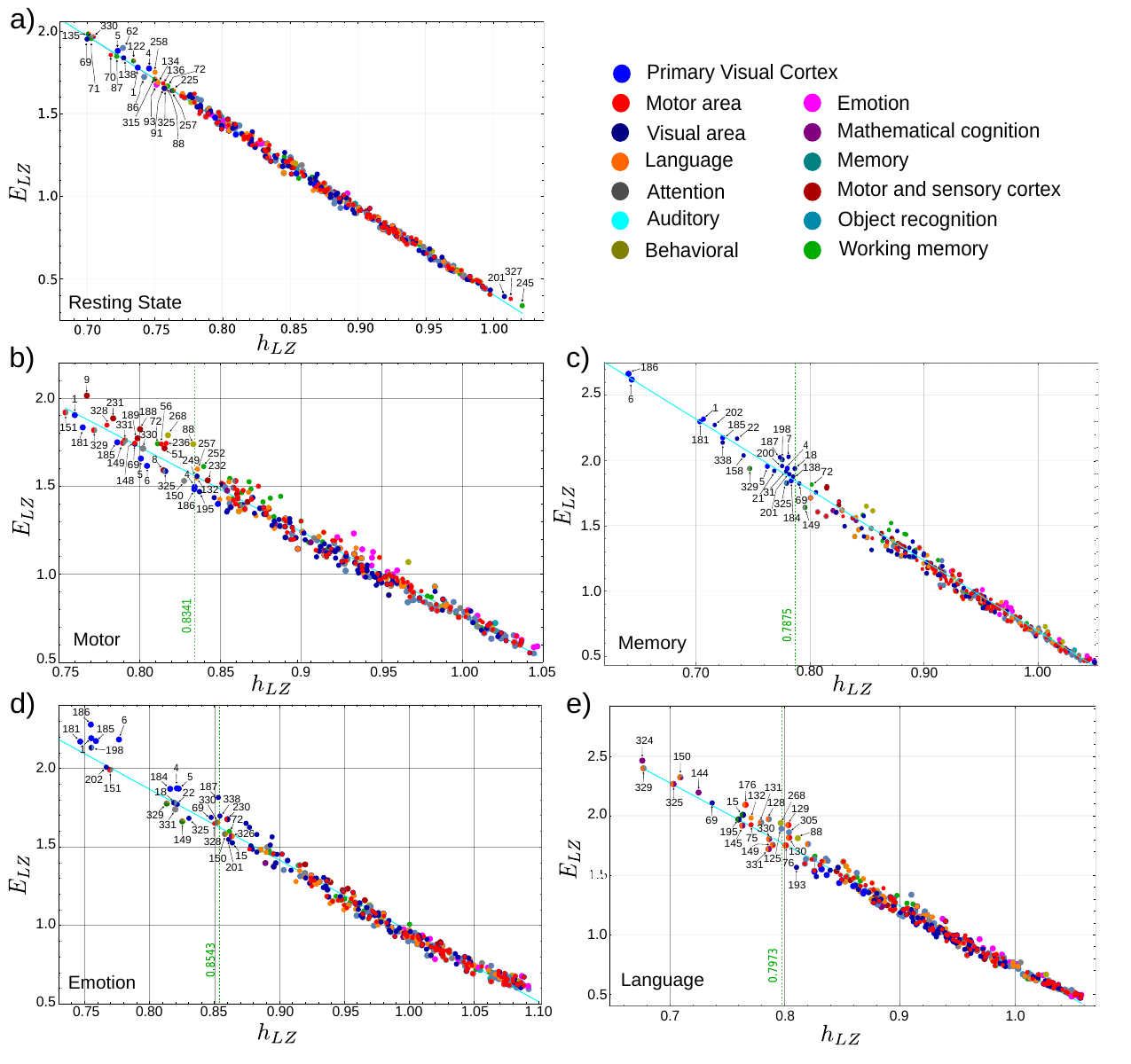}
        
                \caption{\small{ \textbf{Complexity-entropy maps.} LZ-effective complexity ($E_{LZ}$) vs LZ-entropy density ($h_{LZ}$) of all regions during resting state (a) and in the four tasks: b) motor, c) memory, d) emotion, and e) language. The active regions during each task present lower $h_{LZ}$ and higher $E_{LZ}$, which points to less unpredictability and more patterned behavior than the rest of the regions. The results confirm that inactive regions have higher randomness as they do not suppress the noisy background \cite{itoTaskevokedActivityQuenches2020}. The vertical line marks the threshold separating strongly active from low active areas; the criterion for its selection is described in the text and illustrated in the next figure. 
                }}
                \label{fig:4stimulus}              
\end{figure*}

Figure~\ref{fig:4stimulus} shows the complexity-entropy maps for the resting state and the four tasks. Each dot represents a $(h_{LZ}, E_{LZ})$ tuple for one of the $360$ ROIs. All values of entropy density $h_{LZ}$, in all tasks, are above $0.6$ bit/symbol, which implies that none of the regions have a completely predictable or completely unpredictable dynamic. At the same time, the LZ-effective complexity is always above $0.45$ bits/symbol, which means that some pattern formation is present on the fMRI signal. Both behaviors are common conditions for complex dynamics.

\begin{figure*}[ht!]
    \centering
    \includegraphics[width=\textwidth]{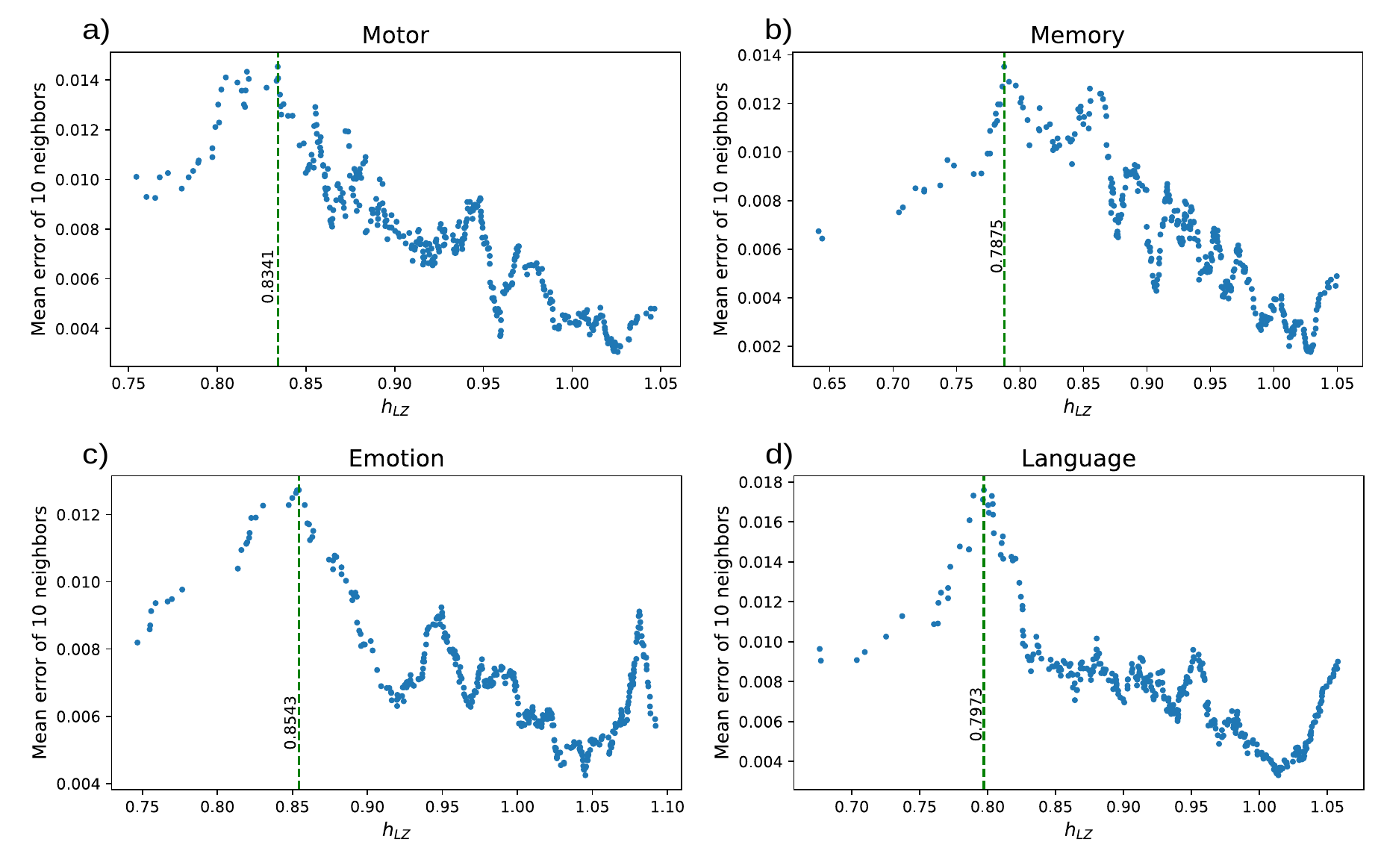}
    \caption{\textbf{Threshold selection criterion for active regions.} For each point in the complexity-entropy map, the mean residual with respect to a linear fit is computed over the point and its ten nearest $h_{LZ}$-neighbors. The threshold is placed at the maximum of this curve, where the scatter significantly deviates from the line on average. This criterion was applied to each task: a) Motor, b) Memory, c) Emotion, and d) Language.}
    \label{fig:crit_act}
\end{figure*}

A strict classification of active and inactive regions in general lacks a clear definition. There is a continuum of points in the complexity-entropy map for the brain regions. This unclear definition of what to consider active or not can lead to subjective criteria, not useful in any analysis; therefore, an objective criterion for segmenting the complexity-entropy map into active and non-active regions is needed. We proceed as follows: for all tasks, there is a linear trend in the data, with a larger dispersion of values for low $h_{LZ}$ and higher $E_{LZ}$. A least squares fit allows us to determine the residuals with respect to the fitted line. From the plot of the residuals, a threshold can be computed by taking for each point, the mean value of its ten $h_{LZ}$ neighbors and using as threshold the maximum value (Figure~\ref{fig:crit_act}). The threshold splits the entropy map into two regions: one with lower (larger) values of $h_{LZ}$ ($E_{LZ}$) and one with larger (lower) values of the same magnitude. According to previous studies \cite{itoTaskevokedActivityQuenches2020}, the inactive regions do not suppress background noise, and therefore, their signal tends to appear more random; we take the lower entropy side (left of the vertical threshold line) to correspond to activated regions. For all tasks, on the active side, there are far fewer regions than on the non-active side.
 
The local averaging is necessary because the raw point-wise residuals are noisy, and locating their maximum directly yields an unstable threshold driven by individual outliers. The neighborhood size $k$ has a practical range of validity: below $k \approx 5$ the averaged curve remains noisy and no clear maximum can be identified, while above $k \approx 20 $the curve becomes nearly monotone in $h_{LZ}$ and the maximum is no longer informative. Within the intermediate range the maximum is well defined and its position varies only mildly with $k$; we use $k = 10$, which lies in the middle of this range. 

Certain regions show specialized activity for different stimuli as reported in previous studies \cite{glasserMultimodalParcellationHuman2016}, and we will follow those findings to understand the complexity-entropy maps.

For the motor task, $8.61\%$ of the regions are active, namely: $1(181)$, $5(185)$, $6(186)$, $8(188)$, $9(189)$, $51(231)$, $56(236)$, $69$, $72$, $88(268)$, $148(328)$, $149(329)$, $150(330)$, $151(331)$,  $325$. The regions with lower entropy density are region PGs (ROI \textbf{151}) and Primary Visual Cortex or V1 (ROI \textbf{1}), very near its hemisphere contrapart ROI \textbf{181}. The first of these areas (ROI \textbf{151}) activates when individuals shift their visuospatial attention from one location to another, specifically in response to biological motion \cite{bakerConnectomicAtlasHuman2018l}, whilst the other areas (ROI \textbf{1} and \textbf{181}) are the most important areas during visual activity \cite{glasserMultimodalParcellationHuman2016}. The area with the largest $E_{LZ}$ value is area 3b, the Primary Sensory Cortex (ROI \textbf{9} and \textbf{189}) which receives information related to bilateral tactile stimulation, this area has been reported to work in conjunction with area 1 (ROI \textbf{231}) primarily involved in processing tactile stimuli \cite{bakerConnectomicAtlasHuman2018d}, which is also within the four regions with the largest LZ-effective complexity values $E_{LZ}$. The motor task asked for a motor reaction to a visual cue.

For the motor task, the Primary and Sensory Motor Cortex regions (ROIs \textbf{188, 9, 189, 51, 231, 52, 232, 53, 233}) tend to lie above the fitted line, except for the Primary Motor Cortex or area 4 of the left hemisphere (ROI \textbf{8}) which lies below. In contrast, the main Visual Areas (ROIs \textbf{1, 181, 4, 184, 5, 185, 6, 186}) all lie below the fitted line.

If we consider the linear regression as signaling the mean performance between $E_{LZ}$ and $h_{LZ}$, then our results can be interpreted that active visual regions underperform in terms of pattern generation for a given entropy density value, as their LZ-effective measure complexity falls below the regression line, and the active Primary Motor and Sensory cortex regions overperform under the same criteria. This motor exercise requires more precisely coordinated movement of fingers, toes, and tongue, while visual attention can allow itself to be more loose as visual cues are given.

Other active areas on the left side of the threshold value are primarily involved in attentional processing and/or motor cue tasks. These include the previously described PGs (ROIs \textbf{151} and \textbf{331}) and  PGi (ROIs \textbf{150} and \textbf{330}), which share many functions. The PFm Complex (ROIs \textbf{149} and \textbf{329}) plays a key role in attentional processing, decision-making, language syntax, and is activated in working memory, motor cues, and gambling tasks. The PF Complex (ROIs \textbf{148} and \textbf{328}) is primarily activated during motor cue and social Theory of Mind (ToM) tasks. Ventral Area 6 (ROIs \textbf{56} and \textbf{236}), part of the ventral secondary motor cortex, is associated with upper body movements (arms, neck, face, and mouth) when electrically stimulated. Additionally, Intra Parietal Area 1 (ROI \textbf{325}), which is also active in motor stimuli, has been consistently identified with interpreting motor cues, listening to stories, and mathematical problem solving, including mental arithmetic tasks, and is more active when processing faces than processing shapes \cite{wuFunctionalHeterogeneityInferior2009, bakerConnectomicAtlasHuman2018l}. 

A similar analysis can be made for the other task stimulus. During the motor, memory, and emotion tasks, subjects relied on visual cues to guide their actions, which corresponds to strong activation in visual regions across all three tasks. While visual areas are highly active in tasks with visual cues, they are less engaged in the language task, where the stimulus is auditory. Yet some differences may be relevant.

In the emotion task, $4.72\%$ of the regions are active, namely: $1(181)$, $4(184)$, $5(185)$, $6(186)$, $18(198)$, $22(202)$, $69$, $149(329)$, $151(331)$,  $187$, $325$, $328$, $330$. The main visual areas overperformed with respect to the regression line, in contrast to the motor task. The emotional exercise involves recognizing faces and their emotional stand. The visual and face-processing regions (V1–V4, FFC, PIT) are more strongly engaged than the attentional regions, with significantly lower $h_{LZ}$ and higher $E_{LZ}$ across subjects (paired across the 153 subjects: $\Delta h_{LZ} = -0.049, p = 1.5\times 10^{-13}, d_z = -0.66; \Delta E_{LZ} = +0.333, p = 6.6\times10^{-16}, d_z = +0.73$; Supplementary Section~4, Table~3). The Fusiform Face Complex or FFC (ROIs \textbf{18} and \textbf{198}), involved in face perception and identification, plays a role in emotional expression recognition due to its high activity in response to both neutral and expressive faces. \cite{chenFunctionallyStructurallyDistinct2023}. For many years, it was believed that damage to this area was the cause of prosopagnosia \cite{steevesFusiformFaceArea2006}. Another paravisual area is the Inferior Temporal Cortex or PIT (ROIs \textbf{22} and \textbf{202}), which is involved in detecting faces and places. Area PIT is associated with processing fundamental aspects of color, such as hue, saturation, and brightness, using input from the eight visual area \cite{bakerConnectomicAtlasHuman2018p} or V8 (ROIs \textbf{7} and \textbf{187}), which is also active. The Inferior Temporal Cortex is a crucial component of the ventral visual pathway, contributing to the recognition of objects, faces, and scenes. It can be observed that this area shows high activity during the emotion tasks.

Several areas of attentional processing are activated during the emotion task, but many of them perform slightly below the regression line, though they are nearly on it. This applies to ROIs \textbf{151, 331, 149, 329, 148, 328, 150} and \textbf{330}, which were previously described in the motor task. In this context, attentional processing is not as critical or strongly linked to task performance as visual areas or face detection regions, which overperformed with respect to the regression line and showed higher activation in general.

Compare the active region in the emotion task with those in the memory task. In the memory exercise, $11.4\%$ of the regions are active, namely: $1(181)$, $4(184)$, $5(185)$, $6(186)$, $7(187)$, $18(198)$, $21(201)$, $22(202)$, $138(318)$, $158(338)$, $200$, $325$, $329$. Here, visual cues are also relevant, yet they are objects and neutral faces that are not supposed to convey emotions. The Fourth Visual Area or V4 (ROIs \textbf{6} and \textbf{186}) are still in the upper left region of the complexity-entropy map, yet the values of $h_{LZ}$ have shifted towards lower values and, correspondingly, the $E_{LZ}$ values have increased. Regions \textbf{185}, \textbf{181}, and \textbf{1}, all visual regions and clustered with ROIs \textbf{6} and \textbf{186} in the emotional task, are still strongly active in the memory exercise but are clearly separated from the last two. In the memory task, both the visual regions and the object recognition areas lie close to the regression line and can not be considered under- or over-performing. The Inferior Temporal Cortex or PIT (ROI \textbf{22} and \textbf{202}) associated with the paravisual area is also strongly activated. The functional specializations referred to here (faces, tools, places, body parts) are those reported for these regions in the parcellation literature; the entropic analysis identifies the regions as active but does not by itself resolve category selective processing.

Area V3CD (ROIs \textbf{158} and \textbf{338}) is located between the early visual cortex and the MT complex. It is activated in fMRI scans during tasks involving tools, relational matching contrast, working memory, emotion, and place recognition.  However, it shows less activation in body-related tasks and exhibits variable activation or deactivation in response to face-shape contrasts. Area V3CD combines detailed information about objects, including contrast and motion edges, to form a complete representation for recognizing objects \cite{bakerConnectomicAtlasHuman2018p}. This area is highly activated, as expected, during the memory task.

The Lateral Occipital Cortex plays a key role in object processing, encoding, and recognition \cite{grill-spectorLateralOccipitalComplex2001, malachObjectrelatedActivityRevealed1995}. The regions belonging to this cortex are Lateral Occipital areas 1, 2, and 3 or LO1 (ROIs \textbf{20} and \textbf{200}), LO2 (ROIs \textbf{21} and \textbf{201}), and LO3 (ROIs \textbf{159} and \textbf{339}). They are adjacent to areas PIT and V3CD, which are strongly activated during memory tasks involving object recognition and are located near the early visual cortex. We observed strong activation of LO1 and LO2 during memory tasks, but not of LO3, which is considered a new area in the Glasser study. LO3 has been shown to be more involved in the processing of the details, motion, and shape information of objects \cite{bakerConnectomicAtlasHuman2018p}.

In addition, the remaining active areas are mostly paravisual regions that have been associated in the literature with object and face recognition; these include ROIs \textbf{18, 198, 7, 187, 325} and \textbf{138}, with the latter associated in prior work with tool and place recognition \cite{bakerConnectomicAtlasHuman2018p, glasserMultimodalParcellationHuman2016}. Additionally, attentional processing areas, such as ROIs \textbf{149} and \textbf{329}, are also active.

In the language exercise, $13.61\%$ of the regions are active, namely: $15(195)$, $69$, $75$, $125$, $128$, $131$, $132$, $144(324)$, $145(325)$, $149(329)$, $150(330)$, $176$, $263$, $268$, $331$. This task interleaves story and math blocks within a single run; since the entropic measures are computed over the whole run, the activation reflects both components and is not separated into story and math specific contributions. No visual cue is given, and accordingly, the visual regions show no strong engagement in the complexity-entropy map, consistent with the HCP group GLM, in which the visual networks are task-negative during the language run (Supplementary Table 6). In contrast, areas associated with math are now strongly stimulated. The language exercise involves auditory cues and some regions of the auditory cortex have become active, such are the cases for the Superior Temporal Sulcus ventral anterior or STSva (ROIs \textbf{176}) and the Auditory Association Cortex (ROI \textbf{128}) which have not been activated in the other task exercise, and are mainly implicated in speech processing, specifically in the story-math contrast and in primary language tasks \cite{glasserMultimodalParcellationHuman2016}. The superior temporal sulcus (STS) is referred to by some authors as the "chameleon" of the human brain due to its involvement in a wide range of functions \cite{heinSuperiorTemporalSulcus2008}; in particular, the anterior region (STSva) is largely dedicated to speech processing.

In the language stimuli, Intra Parietal Areas 1 and 2 (IP1: ROIs \textbf{145} and \textbf{325}, and IP2: ROIs \textbf{144} and \textbf{324}) are strongly active. These are regions involved in mathematical cognition, consistently implicated in mental arithmetic, and their activation here reflects the math blocks of the language run rather than story processing \cite{wuFunctionalHeterogeneityInferior2009, bakerConnectomicAtlasHuman2018l}. IP2 area (ROI \textbf{324}) shows the lowest $h_{LZ}$ value and the strongest $E_{LZ}$ values during the language stimulus but not in the other tasks, consistent with the arithmetic content of this condition. Area IP1 has also been reported to activate when interpreting motor cues and listening to a story, in addition to arithmetic. ROI \textbf{325}, strongly activated in the language task, also shows slight activation in the motor, memory, and emotion tasks, possibly due to its role in interpreting motor cues.

The area PGi (ROIs \textbf{150} and \textbf{330}) shows greater activity when listening to a story, especially compared to answering arithmetic problems, and also exhibits increased activation when processing faces \cite{glasserMultimodalParcellationHuman2016}. Additionally, this region becomes active when individuals shift their visuospatial attention from one location to another \cite{bakerConnectomicAtlasHuman2018l}. These patterns are reflected in the entropic measures for language, where it is strongly activated. Additionally, this area shows strong activation in response to motor stimuli as well. The temporal area 1 anterior or TE1a (ROIs \textbf{132}) is more active in semantic pathways than visual pathways, showing greater activation during language processing tasks \cite{glasserMultimodalParcellationHuman2016} and can be seen as active in the corresponding complexity-entropy map.

In the language exercise, the active regions can not be classified as under- or over-performing as they lie close to the linear regression fit.

\begin{figure}[ht!]
         \centering
             \includegraphics[width=0.8\textwidth]{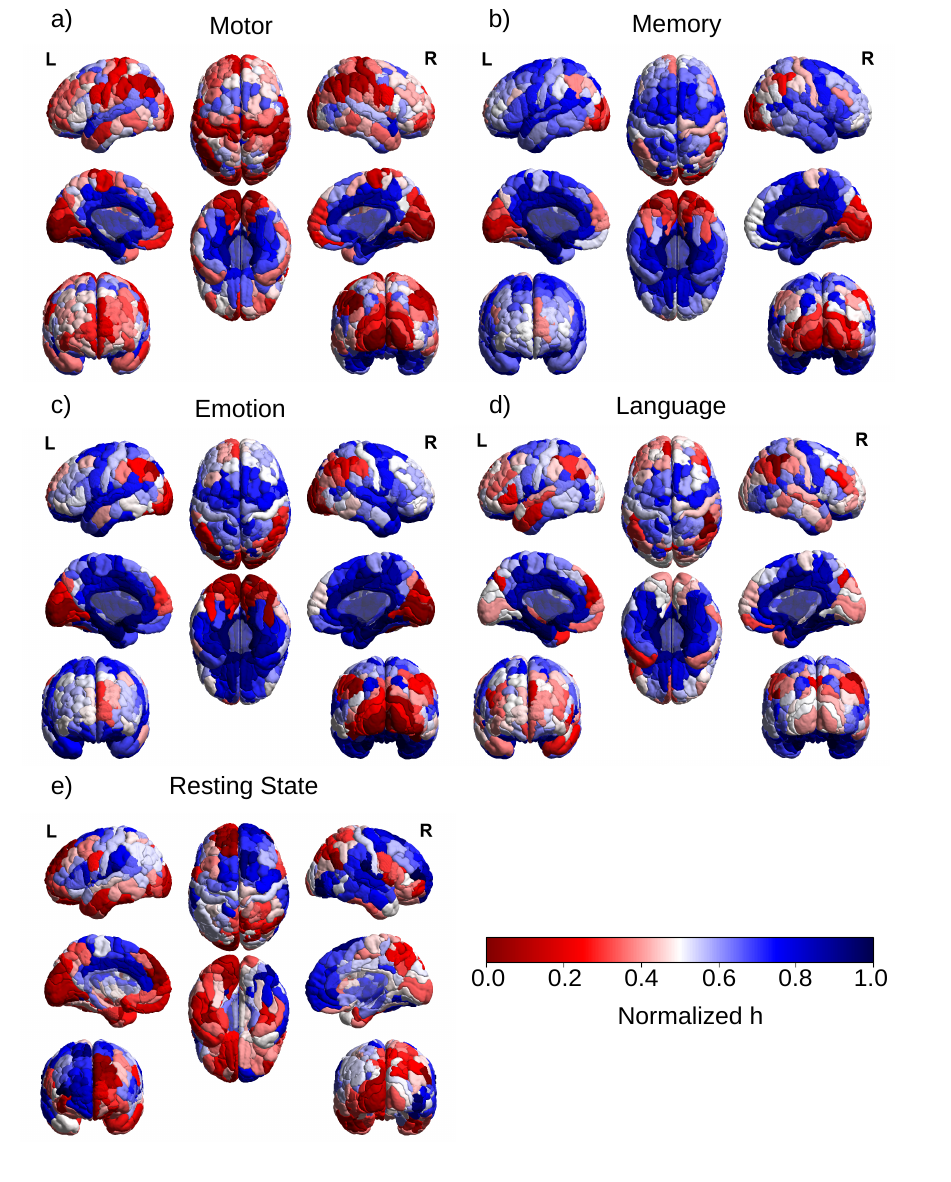}
             \caption{ \textbf{Brain mapping of activity levels.} Level of activation by the normalized entropy density in each stimulus: a) Motor, b) Memory, c) Emotion, d) Language, and e) Resting State. At lower $h_{LZ}$ (red color) the regions are more active than at higher $h_{LZ}$ (blue color). Notice how the visual regions in the occipital area are strongly activated during motor, emotion recognition, and working memory stimuli, which have a visual cue. In contrast, during language stimuli, the visual areas are not active, but the auditory areas are. Observe the strong activation of the motor cortex in the posterior precentral gyrus, immediately anterior to the central sulcus, during the motor task but not in the others. Similarly, note the strong activation of the FFA and PIT -paravisual areas specialized in face and object recognition, respectively- in the memory and emotion tasks, where faces are included in the stimuli; however, these areas are not active in the other tasks. }
             \label{fig:colorBrain}
 \end{figure}

For a better visualization of brain activity using entropy density, Figure \ref{fig:colorBrain} illustrates the activation levels of various brain regions during each stimulus, mapped onto the brain surface. The figure provides a detailed view of how different stimuli engage specific areas, highlighting the spatial distribution of neural activity. As previously discussed, lower values of $h_{LZ}$ (represented in red) correspond to the most active regions involved in task processing, while higher $h_{LZ}$  values (represented in blue) indicate lower activation levels. The exact locations of the discussed regions on the brain surface can be found in the Supplementary material Figure 1.

Notice how the tasks involving a visual cue, such as motor, working memory, and emotion recognition, show strong activation in the main visual areas located in the occipital lobe; this is not observed in the language task, where the stimulus is auditory. Another notable pattern is the strong activation of the motor and sensory cortices, located in the posterior precentral gyrus, just anterior to the central sulcus, during the motor task in which the subject is asked to move their extremities; this activation is not present in the other tasks that are not focused on bodily movement.

On the other hand, we highlight the key role of the Face Fusiform Area (FFA) during the emotion stimulus, where the subject observes faces to recognize emotions. This region is also active during the memory task, although a bit weaker, as the visual attention is distributed across faces, places, tools, and body parts. In contrast, the FFA shows no activation during motor and language tasks, which do not involve facial stimuli.

Paravisual areas involved in contrast, saturation, color detection, and, more generally, in the processing, encoding, and recognition of objects, such as the Inferior Temporal Cortex (PIT), Area PH, Occipital Scene Area (V3CD), and Lateral Occipital Cortex, are strongly activated during the Memory and Emotion stimuli, where subjects are engaged in recognizing and interpreting visual input. These regions are not active during the language and motor tasks, which do not require that level of visual processing.

The supplementary material can be consulted to locate the active regions involved in the language task and to see how they are typically classified in the literature. These include Broca's areas, associated with language comprehension and production; mathematical processing areas, engaged during arithmetic problem-solving; and auditory regions, which are strongly activated during the language task but not in response to the other stimuli.

The attentional processing areas identified in the supplementary material correspond to the previously described regions: PGs (ROIs \textbf{151} and \textbf{331}), PGi (ROIs \textbf{150} and \textbf{330}), PFm Complex (ROIs \textbf{149} and \textbf{329}), and PF Complex (ROIs \textbf{148} and \textbf{328}). These regions exhibit varying levels of activity across different stimuli.

In general, the motor regions are the $51.8\%$, $23.8\%$, $4.34\%$, and $5\%$ of the active regions for the motor, emotion, memory, and language, respectively, being the largest percent for the motor task, as should be expected.  While the visual regions are the $29.6\%$, $57.1\%$, $96.6\%$, and $10\%$ of the active regions for the motor, emotion, memory, and language, respectively, being maximum for the memory task, where it clearly plays a relevant role. In the language task, the language regions account for $65\%$ of the active regions.

The entropic framework presented here is complementary to, rather than competitive with, GLM-based activation analyses and correlation or coherence based connectivity analyses. The set of regions identified as active in the complexity entropy maps largely overlaps with task-relevant areas recoverable by a GLM contrast, providing validation of the entropic measures as reliable indicators of regional engagement. Beyond this overlap, the $(h_{LZ}, E_{LZ})$ map characterizes each region by a position in a two-dimensional plane rather than by a single amplitude estimate, separating regions that share an activation status but differ in the regime of their dynamics, as illustrated by the contrasting position of motor and visual regions above and below the regression line in the motor and emotion tasks. The procedure operates directly on the BOLD time series of each region and requires no design matrix or hemodynamic response model, allowing the same analysis to be applied uniformly to task and resting-state data. 

Inter-subject variability of the entropic measures and the individual versus group reproducibility of the activation pattern were quantified and are reported in Supplementary Section 8 and 9 (Table 7, Figures 9 and 10)

\subsection*{Areas that are identified as active in this study but are not typically considered activated}\label{subsec3.1}

We look into regions found active in the complexity-entropy map, but not usually reported as active for the corresponding task.

Area 9m (ROIs \textbf{69}) is primarily located in the medial frontal lobe, and represents a more recent subdivision of the previously defined area 9 by Petrides and Pandya \cite{petrides1999dorsolateral}, which is a key part of the dorsolateral prefrontal cortex \cite{bakerConnectomicAtlasHuman2018f}. This region exhibits increased activity when monitoring multiple pieces of spatial information. Compared to its superior neighbor, area 9p, area 9m is less deactivated (or more activated) in the language story contrast and displays stronger activation in the TOM-Random contrast \cite{glasserMultimodalParcellationHuman2016}. Although area 9m is not typically associated with language and algorithmic tasks, previous studies have indicated that it is less deactivated during the language story task. Our measures reveal strong activation in the language and motor exercises, along with slight activation in the memory, emotion tasks, potentially due to its role in monitoring multiple pieces of spatial information.

The Parieto-Occipital Sulcus area 2 or POS2 (ROI \textbf{195}) shows greater activation in the motor CUE-AVG contrast and reduced activation in the Face-AVG contrast, while it is more deactivated in the language, math, story, TOM-random, and faces-shapes contrasts, according to the study by Glasser et al. (2016) \cite{glasserMultimodalParcellationHuman2016}. However, in 2022, Matthew F. Glasser observed that this region demonstrated a relative preference for math \cite{assemDomainGeneralCognitiveCore2020}; though they noted that math preferences may be confounded with auditory preferences in their data. The complexity-entropy map found strong activation in the language stimulus, where subjects were challenged with both story and math tasks.

The area 10v (ROIs \textbf{88} and \textbf{268}) is a newly identified ventral subregion of the preexisting area 10, which is one of the decision-making areas of the brain. This region contributes to behavioral decision-making by integrating information from the orbitofrontal cortex and the anterior cingulate cortex \cite{bakerConnectomicAtlasHuman2018f}. In our entropic analysis, this region is highly activated during the motor task and slightly active in response to language stimuli.

Area 10d (ROIs \textbf{72}) is the dorsal subdivision of Area 10 and is involved in episodic and working memory tasks, contributing to abstract cognitive functions \cite{bakerConnectomicAtlasHuman2018b}. This region shows strong activation during the complexity-entropy map of the motor task.

Intra Parietal Areas 1 (IP1: ROIs \textbf{325}) appear active in the emotion task; this area, as explained above, is usually attached to abstract thinking such as mathematical operations (arithmetic tasks), and also when interpreting motor cues and listening to a story \cite{wuFunctionalHeterogeneityInferior2009, bakerConnectomicAtlasHuman2018l}.

Area PGs (ROIs \textbf{331}), which have been reported to be active when individuals shift their visuospatial attention from one location to another, specifically in response to biological motion, are seen in the complexity-entropy map of the language task as activated even though there are no visual clues in this exercise.

It may be surprising that regions of the hippocampus, parahippocampus, and amygdala are not included among active regions during memory stimuli, especially given the attention these areas have received in notable studies of patients with brain lesions, such as that of Henry Molaison \cite{dossani2015legacy}. In Molaison's case, episodic memory was affected due to the removal of these brain areas, while short-term memory remained intact \cite{eichenbaum2011cognitive}. Memory is a highly complex cognitive activity of many types. It is generally divided into long-term memory, which includes semantic, episodic, and procedural memory, among others, and short-term memory, which includes working memory. Working memory is characterized by the manipulation and active response to briefly stored information and is the type of memory of our current stimulus. Also, the amygdala and hippocampus are located beneath the cortical surface and are not directly represented in the Glasser atlas, which primarily focuses on the cerebral cortex \cite{briggsConnectomicAtlasHuman2018}.

\subsection*{The resting state as seen through the entropic magnitudes}\label{subsec3.2}

During resting-state fMRI (rfMRI) scans, participants kept their eyes open, maintaining a relaxed fixation on a bright cross-hair projected onto a dark background in a dark room.

The resting state does not show the dispersion of values with respect to the linear regression fit, the threshold value as calculated for the specific task can not be performed, however, from the grouping of the low $h_{LZ}$ points, 22 regions can be considered in the far left end of the plot as those more active. 

The active regions from the left hemisphere can be grouped by physical continuity into three classes. The region with the lowest entropy density value and larger effective complexity measure is region 135(TF), belonging to the Temporal lobe, which together with its bordering neighboring regions 122(PeEc), 131(TE2a), 136(TE2p), all belonging also to the temporal lobe, and the region 138(PH), belonging to the Occipital Lobe. A second border connected class is formed by regions 69(9m), 70(8BL), 71(9p), 72(10d), 86(9-46d) and 87(9a), all from the Lateral Front Lobe, which are also border connected to the active regions 62(d32), 88(10v), 91(11l) and 93(OFC). The third class of border connected active regions all belong to the Occipital Lobe and include regions 1(V1), 4(V2), and 5(V3). In the right hemisphere, regions 225(7Am), 257(a47v), 258(6r), 315(TF), 330(PGi) appear active but have no border connection between them. The clear contrast between the more region-activated left hemisphere and the less region-activated right hemisphere was evaluated using an asymmetry index. We computed the relative difference between hemisphere pairs for all 180 regions. The calculation involves taking the absolute value of the entropy density difference between both regions in a pair and dividing it by the average value between them. Then the mean value over all relative differences is taken as the asymmetry index. The results corroborate the higher asymmetry between hemispheres in the resting state than in each specific task; the asymmetry index for the resting state was $0.41$ while the same index gave the values $0.079$, $0.078$, $0.11$, and $0.084$ for the motor, emotion, language, and memory experiments, respectively. All indices were computed on data processed with the HCP minimal preprocessing pipeline, including motion correction and artifact and physiological-noise removal applied identically to the task and resting-state runs; because the index is a relative measure between homotopic pairs, condition-wide nuisances that shift both homologs together cancel in their difference and do not inflate the index.

From a functional perspective, the most represented functions of the active regions in the resting state are visual tasks with 11 regions, 8 regions with working memory, 8 regions with language processing, and 7 regions with movement. It must be recalled that the resting experiment was performed with eyes open, focusing on a geometric object.

The largest $E_{LZ}$ region in the resting state had been associated with memory (also face recognition, visual area, and language processing). The next region in terms of $E_{LZ}$ value is 330(PGi) in the right hemisphere, related to language processing, attention, and mathematical tasks. These two regions are followed by the 69(9m) and 71(9p) regions, the first associated with the visual area and spatial information, and the latter with working memory. On the other side, the least activated regions, those with the largest $h_{LZ}$ value, and the smaller $E_{LZ}$ values are regions 201(LO2), 245(10r), and 327(PFop) on the right hemisphere.

\subsection*{Validation against established methods.}

To quantify the correspondence between the entropy-derived active regions and established task systems, we performed three independent validations.

We first assessed whether the active set of each task is preferentially associated in the cortical networks of the Cole–Anticevic Brain-wide Network Partition, which assigns each of the 360 Glasser regions to one of twelve functional networks \cite{ji2019mapping} (version 1.1.6). 

Fold enrichment of a network is the fraction of the active regions that belong to that network divided by the fraction of all cortical regions that belong to it : $fold = (k/n) / (K/N)$, where k is the number of active regions in the network, n the total number of active regions, K the number of regions the network contains, and N = 360 the total number of regions. A value of 1 indicates the network holds as many active regions as expected if the active set were distributed at random across the cortex; values above 1 indicate over-representation. Significance was assessed per task with a hypergeometric test, which determines the probability of finding a specific number of successes in a sample drawn from a finite population and can be found in Table~4 of the supplementary material.

\begin{figure}[ht!]
    \centering 
    \includegraphics[width=\textwidth]{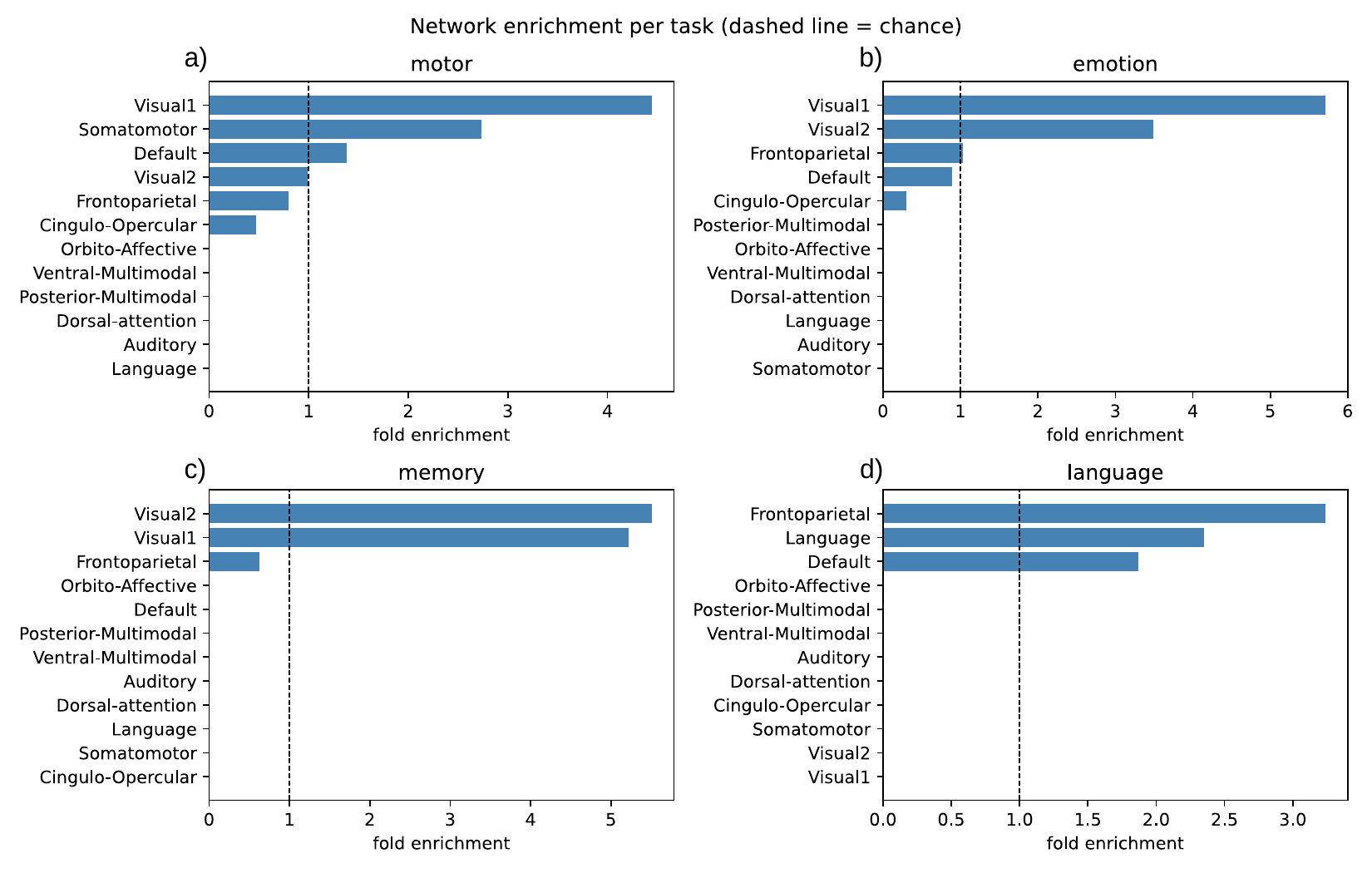}
    \caption{\textbf{Network enrichment of the entropy-active regions.} For each task, fold enrichment of the active set in the twelve Cole–Anticevic networks. Each task's active set is concentrated in its expected network. The dashed lines are the values of fold enrichment by chance.}   
    \label{fig:network_enrichment}
\end{figure}

For every task the active set is concentrated in the expected network (Figure \ref{fig:network_enrichment}): working memory in the visual networks ($5.5$-fold enrichment, $p = 5.7\times10^{-14}$), emotion in the visual networks ($3.5$-fold, $p = 3.6\times10^{-5}$), language in the frontoparietal ($3.2$-fold, $p = 4.9\times10^{-4}$) and language networks, and motor in the somatomotor network ($2.7$-fold, $p = 0.004$), the motor set being additionally distributed over the visual and cingulo-opercular networks engaged by the visual cue. The full network by task enrichment is given in a heatmap (Supplementary Figure 7) and a table (Supplementary Table 4).

Second, we compared the regions identified by the entropic framework with meta-analytic activation maps from Neurosynth (http://neurosynth.org), which aggregates coordinates from 14 371 fMRI studies and produces term-specific association maps via a chi-squared test against studies that do not mention the term. For each task we selected terms matching its cognitive components: \emph{motor}, \emph{finger tapping}, \emph{foot} and \emph{visual} for the motor task; \emph{working memory} and \emph{visual} for the memory task; \emph{emotion}, \emph{emotional faces} and \emph{visual} for the emotion task; and \emph{language}, \emph{arithmetic} and \emph{listening} for the language task. Maps were generated using NiMARE~\cite{Salo2023, salo_taylor_2022_5826281}.

\begin{figure}[ht!]
    \centering 
    \includegraphics[width=\textwidth]{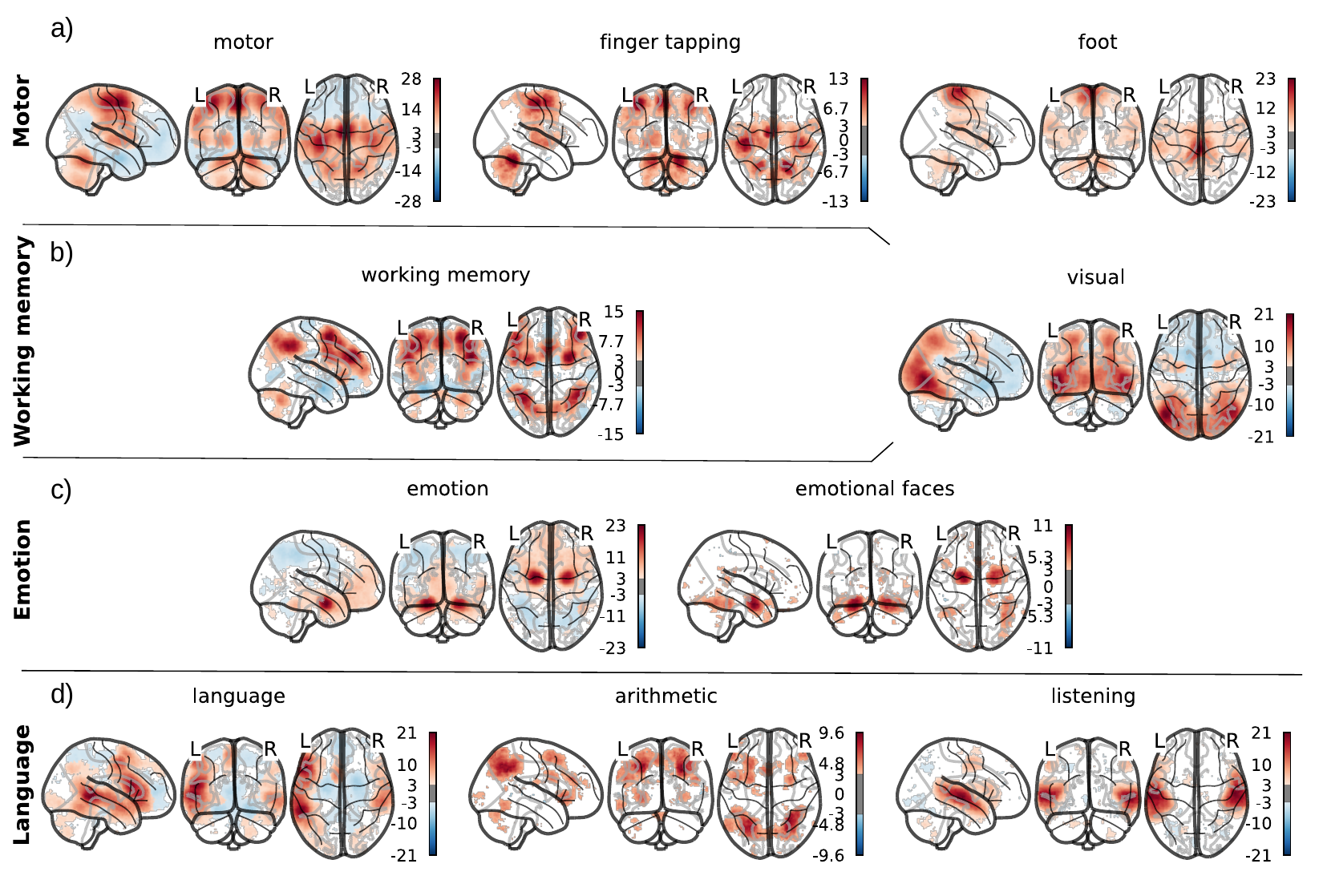}
    \caption{\textbf{Neurosynth meta-analytic association maps for the four HCP tasks.} Each panel shows glass-brain renderings of the Neurosynth association z-maps (thresholded at $z > 3)$ for the terms used in the validation analysis. (a) Motor task: motor, finger tapping, foot, visual. (b) Working memory task: working memory, visual. (c) Emotion task: emotion, emotional faces, visual. (d) Language task: language, arithmetic, listening. These maps are to be compared with the entropic activation maps for the same tasks shown in Figure \ref{fig:colorBrain}. Regions of general engagement, primary visual cortex (engaged by any visually-cued task) and the lateral and inferior parietal cortex (engaged by attentional demands), are prominent in the entropic maps but largely absent from the term-specific Neurosynth maps, because the Neurosynth association test isolates regions specifically associated with each cognitive term and removes regions activated generically across most fMRI paradigms. The visual areas identified by Neurosynth correspond closely to the active regions observed in the entropic paradigm for visually cued tasks. Neurosynth maps are volumetric and include subcortical structures, most visible as the central activations in the emotion and emotional faces maps (amygdala and hippocampus), which are not represented in the cortical Glasser parcellation used by the entropic framework.}   
    \label{fig:neurosynth_comparison}
\end{figure}

Figure~\ref{fig:neurosynth_comparison} shows the Neurosynth maps for the ten terms. The qualitative comparison with the entropic activation maps in Figure~\ref{fig:colorBrain} reveals several similarities. Regions activated by general engagement, primary and early visual cortex in the cued tasks, and lateral and inferior parietal cortex under attentional demands, are prominent in the entropic maps but absent from the term-specific Neurosynth maps, except for the keyword visual, which is why we included visual in the analysis. This follows from the Neurosynth association test: regions activated across most fMRI experiments regardless of task contribute equally to studies that mention any term and to those that do not, and are removed by the chi-squared statistic. Second, regions where the literature is consistent for each term are emphasized in the Neurosynth maps: supplementary motor area and premotor cortex for motor, dorsolateral prefrontal cortex and intraparietal sulcus for working memory, inferior frontal and superior temporal regions for language (strongly on the left). These regions are also recovered by the entropic framework.

The partial agreement is explained as follows. Neurosynth maps reflect the cumulative published literature for each term and may emphasize different sub-regions than a specific task design. The motor literature is dominated by studies of motor planning and sequencing, weighting the motor map toward supplementary and premotor cortex; the HCP motor task is simple finger, foot and tongue movement, engaging primary motor and sensory cortex more strongly. The agreement between the two maps captures the shared core of each cognitive network rather than the full set of regions identified by either approach. The Glasser parcellation used here is restricted to the cerebral cortex, while Neurosynth maps are volumetric and include subcortical structures. For terms whose literature most strongly implicates subcortical regions, the comparison is confined to the cortical component of the Neurosynth map. This is particularly relevant for the emotion task: the most consistently reported regions in the emotion literature are the amygdala and hippocampus, which are visible as the dominant central activations in the Neurosynth emotion and emotional faces maps, but are not represented in our cortical parcellation. The entropic framework instead tracks the visual face-processing engagement of the HCP emotion paradigm, the dominant cortical demand of the task, which matches the visual Neurosynth map more closely than the emotion term itself.

As a third validation, we compared the entropy maps with the HCP group average task GLM activation maps (Cohen's d, 997 subjects, HCP S1200), parcellated to the same atlas. For each task the GLM reference was the per-region maximum of the task versus baseline condition contrasts (movement and cue for motor, the two memory loads for working memory, faces and shapes for emotion, story and math for language), so that a region counts as active if it responds in any condition of the task. Agreement was measured by the Spearman correlation between the entropic activation $(-h_{LZ})$ and the GLM effect size.

The entropic activation reflects the temporal structure of a region's signal over the whole run, whereas the GLM measures response amplitude against a task-timed model; the two are different quantities. Across all 360 regions the correlation is therefore positive but weak ($\rho \in [0.17,0.45]$, Table~5), because around 300 regions that neither method identifies as task-relevant contribute only noise and dominate the count. We restricted the correlation to each task's functional network, as defined by the Cole–Anticevic partition, removing this dilution and isolating the regions where both methods carry signal. Within the somatomotor and visual networks for motor, and the visual networks for emotion and working memory, the entropic activation correlates strongly with the GLM effect size $(\rho = 0.47, p = 7\times10^{-7}; \rho = 0.59, p = 8.6\times10^{-7}; \rho = 0.67, p = 4.6\times10^{-9}$; all FDR-corrected across tasks, Table~5). The entropic measure thus recovers the same graded pattern of regional engagement as the GLM wherever the task drives a positive amplitude response.

The language task is the exception and is informative. Within its full network (Language, Frontoparietal and Auditory) the correlation is null $(\rho = -0.06, p = 0.57)$, but within the Auditory network alone it is strong $(\rho = 0.66, p = 7\times10^{-3})$. The difference is explained by the sign of the GLM response per network (Supplementary Table~6): the GLM activates the Auditory and Language networks (mean Cohen's $d = +0.72$ and $+0.40; 87\%$ and $83\%$ of regions positive) but deactivates the Frontoparietal network (mean $d = -0.08; 42\%$ positive). The frontoparietal regions are engaged by the arithmetic blocks embedded in the language run, but relative to the run's own baseline they fall below it, so the GLM represents them as task-negative. The entropic measure identifies these regions as active because it quantifies the structure of their dynamics over the whole run, without reference to a baseline. The null language correlation across the full network is therefore not a disagreement about which regions are engaged, but a consequence of the GLM measuring amplitude relative to a baseline while the entropic measure does not; the frontoparietal and language engagement that the GLM cannot represent is recovered by the network-enrichment and Neurosynth validations above.

The three references give a consistent result: the regions identified by the entropic framework are the functional systems engaged by each task.

\subsection*{Symmetry in the brain through subjects during tasks and asymmetry during Resting State}

\begin{figure*}[ht!]
        \centering
            \includegraphics[width=0.7\textwidth]{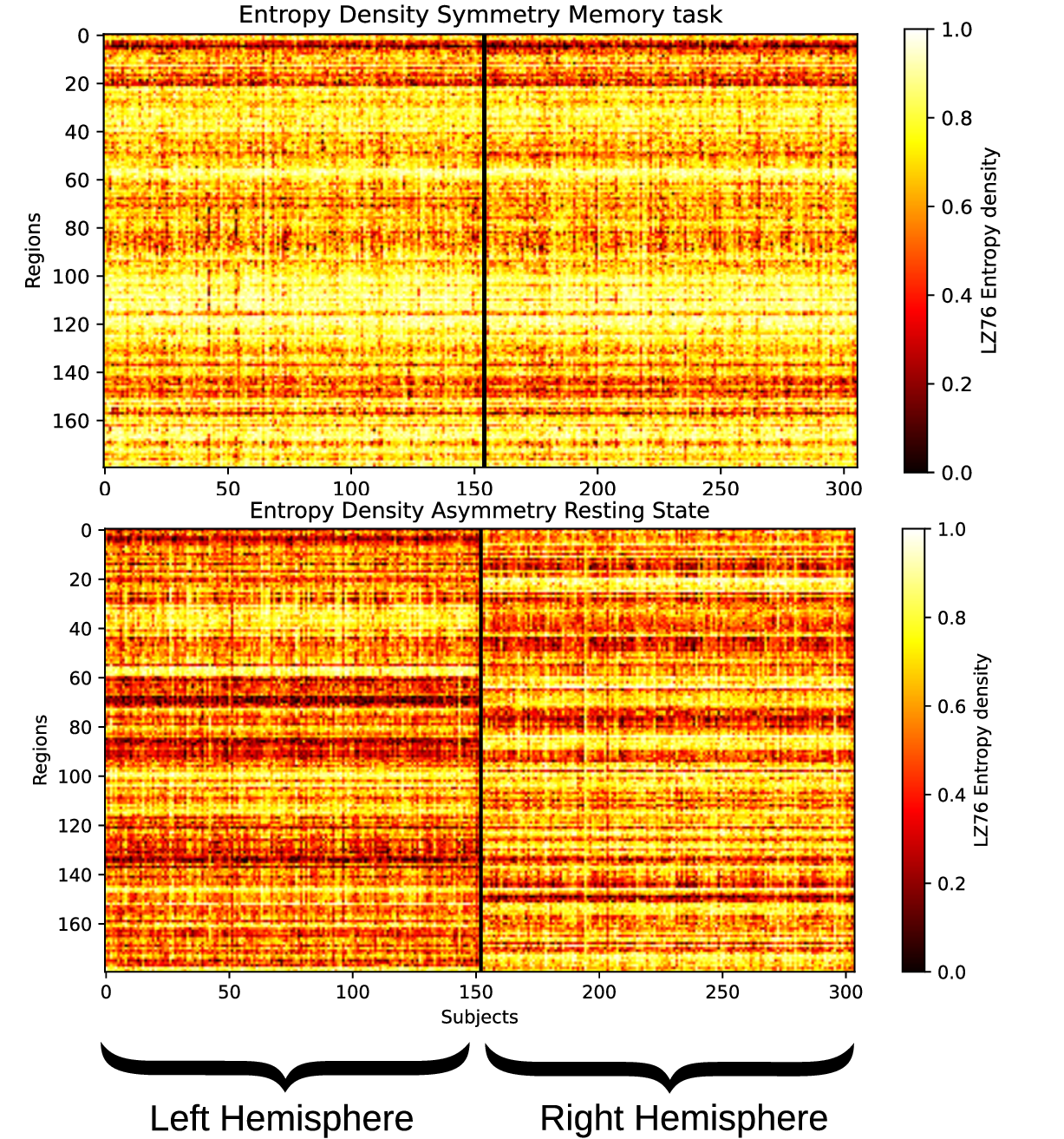}

            \caption{ \textbf{Symmetry and asymmetry across different tasks.} Both figures show the entropy density levels of each region across all subjects (N = 153). The left side displays regions in the left hemisphere (from 1 to 180), while the right side corresponds to the homologous regions in the right hemisphere (from 181 to 360). During task performance, a clear symmetry is observed between hemispheres, with homologous regions exhibiting similar activation levels. In contrast, this symmetry breaks down during the resting state, consistent with previous findings. \cite{raemaekersKnowingLeftRight2018}}
            \label{fig:symmetry_asymmetry}
\end{figure*}

We observe a striking contrast between task-driven and resting-state brain dynamics, as shown in Figure \ref{fig:symmetry_asymmetry} (memory and resting) and further supported by the supplementary figure~6 for the other stimuli (language, motor, and emotion). Specifically, all stimulus conditions exhibit a high degree of interhemispheric symmetry: homotopic regions across the left and right hemispheres display comparable levels of entropy density, suggesting coordinated bilateral activation. However, during the resting state, this symmetry is notably disrupted; entropy density diverges sharply between hemispheres, revealing a strong and widespread asymmetry. This breakdown of symmetry during rest is consistent with previous studies \cite{raemaekersKnowingLeftRight2018, wangBrainAsymmetryNovel2023, saengerHemisphericAsymmetriesFunctional2012} and highlights how, in the absence of external tasks, brain activity becomes more spontaneous and uneven between hemispheres.

\subsection*{Functional brain connections}

\begin{figure*}[ht!]
        \centering
 	\includegraphics[width=\textwidth]{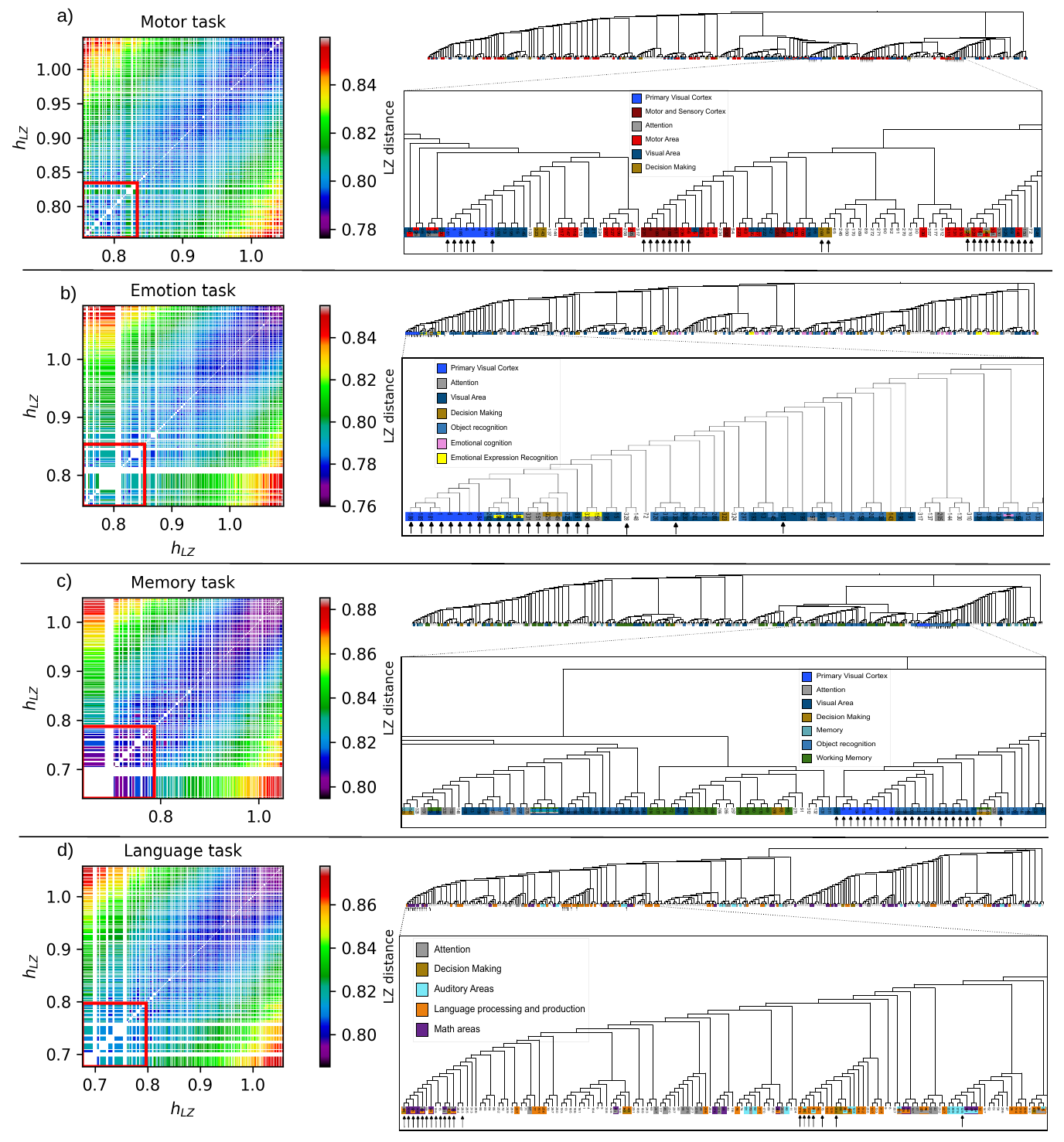}
    \caption{\small{ \textbf{Distance analysis.} The Lempel Ziv distance was calculated between all brain regions for each task: a) Motor, b) Emotion, c) Memory, and d) Language; resulting in the distance matrix shown on the \textbf{left} of each panel. The matrix was rearranged based on activation levels, with lower h values indicating more active regions and higher h values indicating less active ones. The red square highlights the most active regions. As observed, active areas exhibit the greatest distance from non-active areas, while the non-active regions cluster closely together. This suggests that, despite appearing more random, they share underlying dynamics, potentially driven by a rewiring mechanism shaping their behavior. On the \textbf{right} side of each panel, a dendrogram was constructed from the distance matrix to illustrate hierarchical connections. Arrows point to active regions, showing clusters of visual, motor, and attentional processing regions, with strong associations among active regions. }}\label{fig:dist}
\end{figure*}

To analyze the functional connections between brain regions, the LZ-distance was computed from the fMRI time series. Unlike Hamming-type distances, which assess similarity based on point-by-point matches, information distances aim to characterize the relationship between two time series via their normalized mutual information \cite{li04}. In the case of the Lempel-Ziv estimate, this reflects the redundancy of patterns shared by the sequences. This redundancy-based measure is time-independent and does not account for when the related patterns occur in each series. While it is possible to assign a time arrow to information distances, this was not done in the present analysis and will be explored in future work. From the pairwise information distance between all 360 ROIs a distance matrix was computed. We plotted the distance map, sorting both axes by increasing $h_{LZ}$ values. This matrix was then used to construct a dendrogram of the regions, enabling the definition of a hierarchical structure of functional relationships. The dendrogram was built using the Neighbor-Joining clustering method. Distance maps and dendrograms for all four tasks are shown in Figure \ref{fig:dist}. If the reader wants to look closer at the dendrograms, they are provided separately in the Supplementary material Section~10.

\textbf{Motor Task.} Figure~\ref{fig:dist} a) We begin our analysis with the distance map; the larger the difference between the $h_{LZ}$ value of two regions, the larger the distance between them. This trend is valid for both active and non-active regions. In general, the blue area color along the main diagonal shows that similar $h_{LZ}$ valued regions remain close in the sense of pattern redundancy, but this diagonal area does not have the same width for all entropy density values, and it reduces its width notably when entering the active regions portion of the distance map. The opposite happens for the less active regions portion of the same plot.

The first thing that becomes relevant in the dendrogram is that active regions are found clustered together at lower levels of the tree, pointing to shared information between them. Furthermore, regions of similar functionality also appeared clustered in the lower levels of the hierarchy. Both trends are clear, for example, for regions  1(V1), 181(V1), 5(V3), 185(V3), 6(V4), 186(V4), all of them in the occipital lobe and associated with visual tasks, which appear clustered together and also with the other visual area 4(V2), 184(V2), which although not activated according to the criteria used, has an entropy density of $0.83$, quite close to the activation threshold value. A similar example are active regions 8(4), 188(4), 9(3b), 189(3b), 51(1), 231(1), 56(6v), 236(6v) all clustered together at low levels of the dendrogram and the first six regions associated with motor and sensory activity, while the last two belong to the premotor areas. A third more functionally heterogeneous group was also found comprising regions 6(V4), 72(10d), 148(PF), 328(PF),  149(PFm), 329(PFm), 150(PGi), 330(PGi), 151(PGs), 331(PGs), 325(IP1); the last nine regions belong to the lateral parietal lobe, while the first at the occipital lobe and the second at the lateral frontal lobe. This cluster comprises regions with visual, motor, and attention functionalities.

\textbf{Emotion Task. }Figure~\ref{fig:dist} b) A similar behavior was found in the emotional task; the distance map again shows that the larger the difference between the $h_{LZ}$ value of two regions, the larger the distance between them. The dendrogram also shows the clustering of active regions. Now the active visual regions 1(V1), 4(V2), 5(V3), 6(V4), and their opposite hemisphere pair appear clustered at the leaf level. The same goes for regions 18(FFC) and 198(FFC) associated with emotional expression recognition, clustered together with regions 22(PIT) and 202(PIT); the four regions are also associated with object recognition and visual functions. The ROI's 149(PFm), 151(PGs), and there opposite hemisphere pair, together with region 325(IP1) make another cluster related to attention functions. Regions 150(PGi) and 330(PGi) are also mentioned regarding emotional expression recognition. All these clusters are related at higher levels, forming a larger cluster, which did not happen in the motor task, where at least three separate clusters were found.

\textbf{Memory Task.} Figure~\ref{fig:dist} c) The memory task also has a large and unique group of active regions, mostly related to visual functions and object recognition. All active Primary Visual Cortex regions appear together, regions 1(V1), 4(V2), 5(V3), 6(V4) and their opposite hemisphere pairs; at a higher hierarchy level they are clustered with regions 7(V8) and 187, 18(FFC) and 198, 21(LO2) and 201, 22(PIT) and 202, 138(PH) and 318, 158(V3CD) and 338(33pr), 200(LO1) all related to object recognition and visual area.

\textbf{Language Task.} Figure~\ref{fig:dist} d) Two well defined clusters of active regions can be recognized in the dendrogram. The first cluster with ROI's 15(POS2), 69(9m), 144(IP2), 145(IP1), 149(PFm), 150(PGi) and their opposite hemisphere pairs. All these regions have been reported to be associated with Language and mathematical tasks, as well as attention. The second cluster comprises regions active 128(STSda), 131(TGd), 132(TE1a), and 176(STSva), all associated with language processing and production, together with region 75(45).

Across all tasks, the dendrograms show that active regions cluster by functional specialization: regions performing related computations share more pattern redundancy and are therefore closer in the LZ-distance. The non-active regions, despite appearing more random in the complexity-entropy maps, also cluster tightly together, suggesting that they share common background dynamics.
 
As a representative case in the Memory task, to visualize the strongest functional connections, a threshold was applied to the distance matrix, retaining only the pairs with the shortest LZ-distances. The threshold was set at the 5th percentile of all pairwise distance values, preserving the top $5\%$ of strongest connections. The retained connections were represented as a graph, where each node is one ROI and each edge indicates a distance below the threshold. Node size was scaled by the inverse of $h_{LZ}$: larger nodes correspond to more active regions.
 
\begin{figure*}[t]
    \centering
    \includegraphics[width=\textwidth]{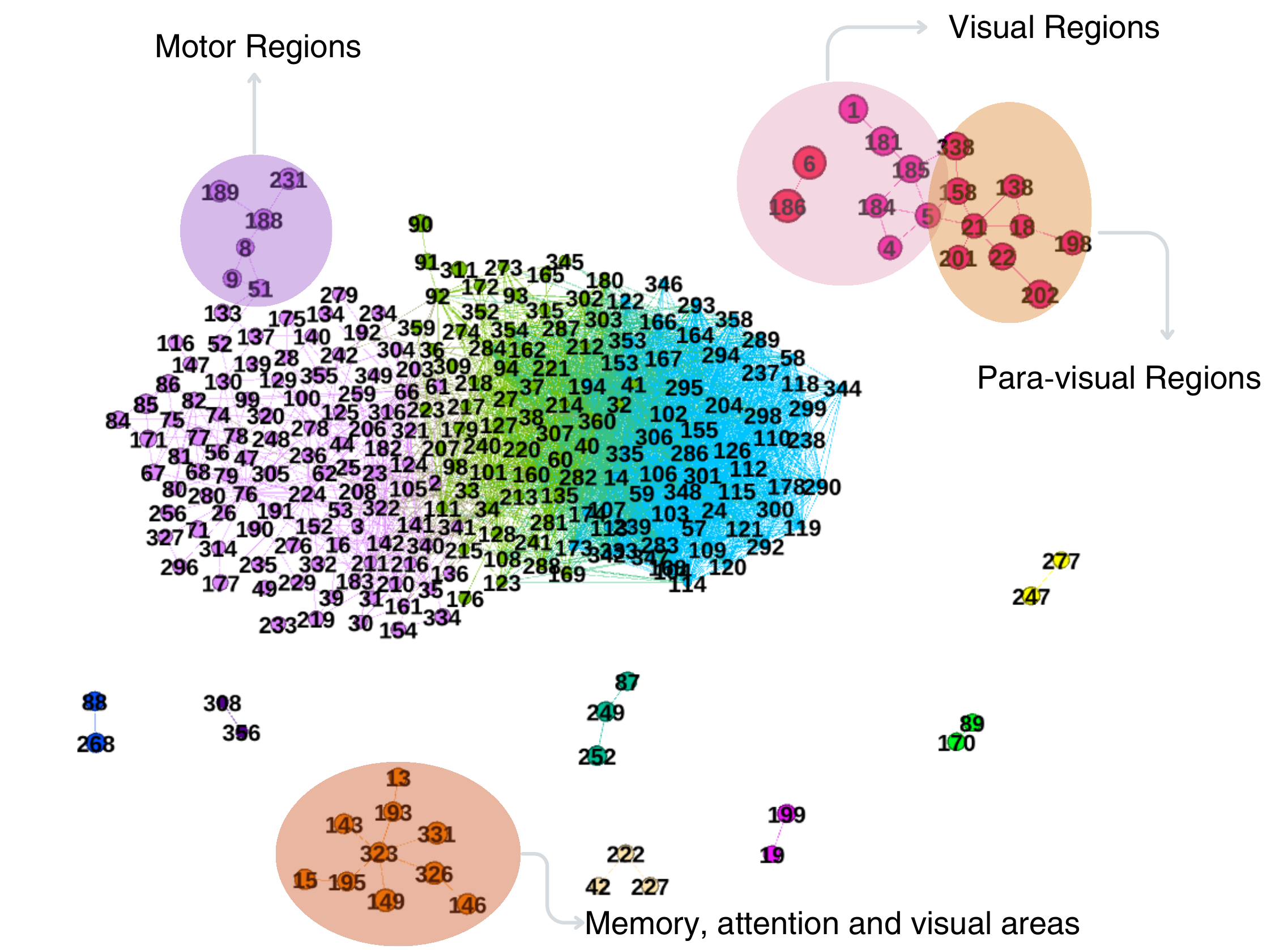}
    \caption{\textbf{Connectivity graph (memory task, 5th-percentile threshold).} Nodes represent ROIs, with size proportional to activation level (inverse $h_{LZ}$). Edges connect pairs whose LZ-distance falls below the 5th percentile. Clusters of visual/paravisual regions, motor regions, and attentional/memory regions emerge at the periphery, while a large central cluster of low-activity regions maintains dense interconnectivity.}
    \label{fig:graph}
\end{figure*}

Figure~\ref{fig:graph} shows the resulting graph for the memory task. The graph exhibits a clear modular structure. Peripheral clusters correspond to functionally specialized groups: visual and paravisual regions form one cluster, motor regions another, and attentional and memory-related regions a third. A large, densely connected central cluster contains the remaining low-activity regions. These regions are strongly connected to one another but have fewer connections to the peripheral, task-active clusters.
 
This structure admits a straightforward interpretation. The central cluster represents a baseline connectivity backbone shared across conditions: non-active regions maintain dense mutual connections regardless of the task. The peripheral clusters represent task-specific subnetworks that become informationally distinct from the backbone during task performance. The active regions are farther from the central cluster and less connected to it, consistent with the observation from the distance matrix that task-driven activity produces specialized patterns that differ from the common background shared by non-active regions.

The connectivity layer, built from the LZ-distance, differs from correlation and coherence in that it is non-linear, time-lag independent, and based on shared pattern emergence rather than on linear co-variation; the modular organization recovered in the connectivity graphs is therefore not reducible to a linear functional connectivity matrix

\section*{Conclusion}\label{sec5}

By applying entropy-based measures, we provide a robust, model-free framework to capture the intricate dynamics of brain activity, with a particular focus on detecting activation patterns through complexity metrics, which are well suited for exploratory analysis and functional connectivity. Our findings show that entropy related measures identify the level of activation capturing non-linear neural dynamics often missed by traditional linear methods. But the metrics offer a richer characterization of the behaviours beyond simply segmenting in active or non-active regions, the $(h_{LZ}, E_{LZ})$ map locates each region along a continuum of complexity and so points directly to its information-processing regime. The consistency and robustness of the analysis across motor, working memory, emotion recognition, and language tasks have been shown.

Since the method does not rely on preconceived models, assumptions about the data, or parameter tuning (beyond the fixed choice of discretization and the activation threshold criterion, neither of which is optimized to the data), it is well suited for exploratory analysis. Using Lempel-Ziv complexity and entropy measures enables the identification of active regions based on the structure and irregularity of their neural dynamics, including regions that are not typically reported under the corresponding task but emerge here through their distinctive complexity profiles. These regions may have been overlooked by traditional linear or model-based approaches under similar stimuli, yet they were relevant for specific tasks related to those excitations.

The Lempel-Ziv distance further allows the framework to move from regional characterization to functional connectivity. Distance matrices, dendrograms, and connectivity graphs derived from the pairwise $d_{LZ}$ reveal a consistent organization across tasks: active regions cluster by functional specialization at low levels of the hierarchy, while non-active regions form a densely interconnected central component. The connectivity graphs show that task-driven activity produces peripheral, functionally specialized subnetworks that become informationally distinct from a common background shared by non-active regions. This dual view, regional engagement from the complexity entropy map and inter-regional structure from the distance based graph, is obtained without any model assumption or parameter tuning beyond the choice of discretization.

A natural extension of the present framework is the introduction of directionality to analyze the effective connectivity, which would allow the recovery of asymmetric, potentially causal relationships between regions; this development will be reported in the future. Together, the results presented here highlight the potential of entropic measures, and Lempel-Ziv estimators in particular, to quantify information transmission between brain regions during task engagement, and to serve as a unified tool for analyzing brain activation and functional connectivity in both research and clinical contexts.

\section*{Acknowledgements}

CITMA is acknowledged for financial support under the project CARDENT, grant PN223LH010-053. Kárel García and Roberto Bernal Arencibia are acknowledged for valuable discussions. The University of Havana and Max Planck Institute for the Physics of Complex Systems are acknowledged for their computer support and working environment. JS Armand Eyebe Fouda is acknowledged for valuable discussions, particularly regarding the creation of Figure \ref{fig:symmetry_asymmetry}.

\section*{Supplementary materials}

\subsection*{Supplementary information 1: Location of key regions on the brain surface}

To better understand the Figure 5 (Main text), we include the following image to indicate the locations of the referenced regions on the brain surface.

\begin{figure*}[ht!]
        \centering
            \includegraphics[width=0.8\textwidth]{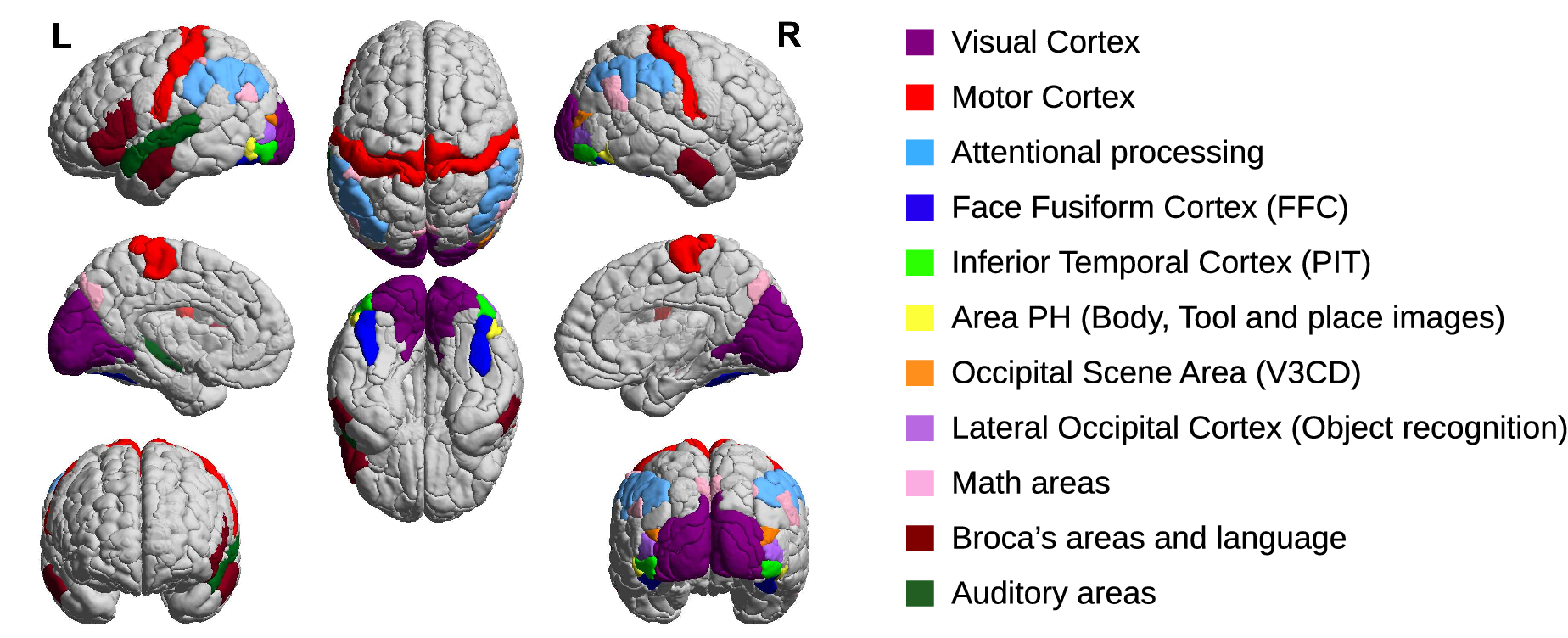}

            \caption{ \textbf{Brain regions location.} Visualization of the brain surface with labeled regions of interest (ROIs) referenced in Figure 5 (Main text)}
            \label{fig:brain_location}
\end{figure*}

\newpage

\subsection*{Supplementary information 2: Spectral entropy versus LZ-entropy density.}

Each point represents one ROI. Spectral entropy $SE$ is the Shannon entropy of the normalized power spectrum of the continuous BOLD signal, computed without any binarization; $h_{LZ}$ is computed on the mean-thresholded binary sequence. The two measures are linearly related across all 360 regions in every condition, confirming that the binarization step preserves the dynamical content relevant to the entropic characterization. A detailed frequency-domain analysis based on $SE$ will be reported separately.

\begin{figure}[ht!]
    \centering
    \includegraphics[width=\textwidth]{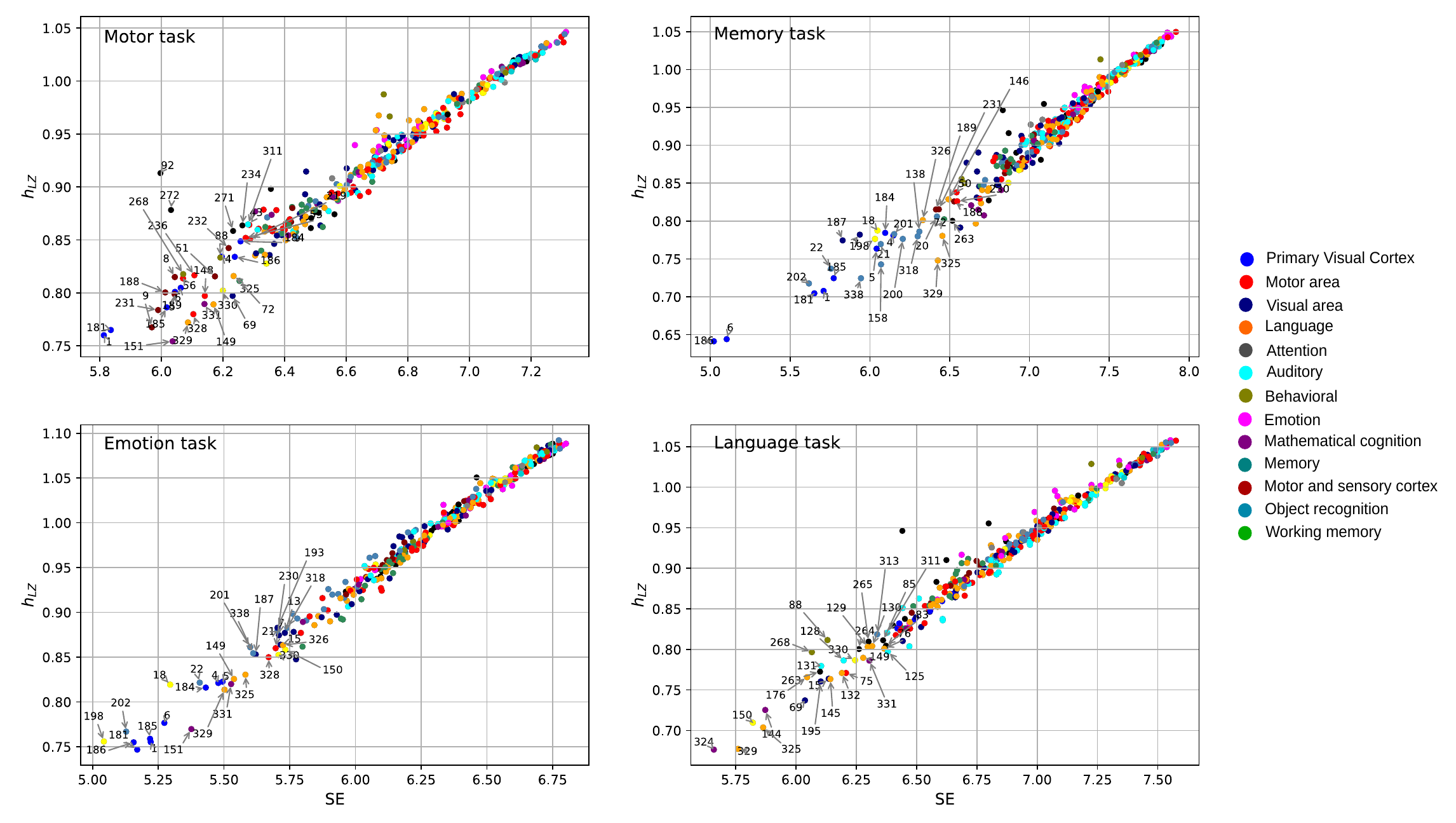}
    \caption{\textbf{Spectral entropy vs.\ LZ-entropy density.} Each point represents one ROI. The linear relationship between $SE$ (computed on the continuous BOLD signal) and $h_{LZ}$ (computed on the binarized signal) is consistent across all tasks, confirming that binarization preserves the dynamical content relevant to the entropic characterization.}
    \label{fig:SE_vs_hLZ}
\end{figure}

\newpage

\subsection*{Supplementary information 3: Robustness of the classification to the discretization.}

To verify that the activation classification and the over/under-performance pattern do not depend on the use of the mean as the binarization threshold, we repeated the full analysis using the median of each region's BOLD signal. As with the mean, $h_{LZ}$ and $E_LZ$ were computed for every region in each subject and then averaged across subjects, and the activation threshold was set with the same residual criterion described in the main text.

\begin{figure}[ht!]
    \centering
    \includegraphics[width=\textwidth]{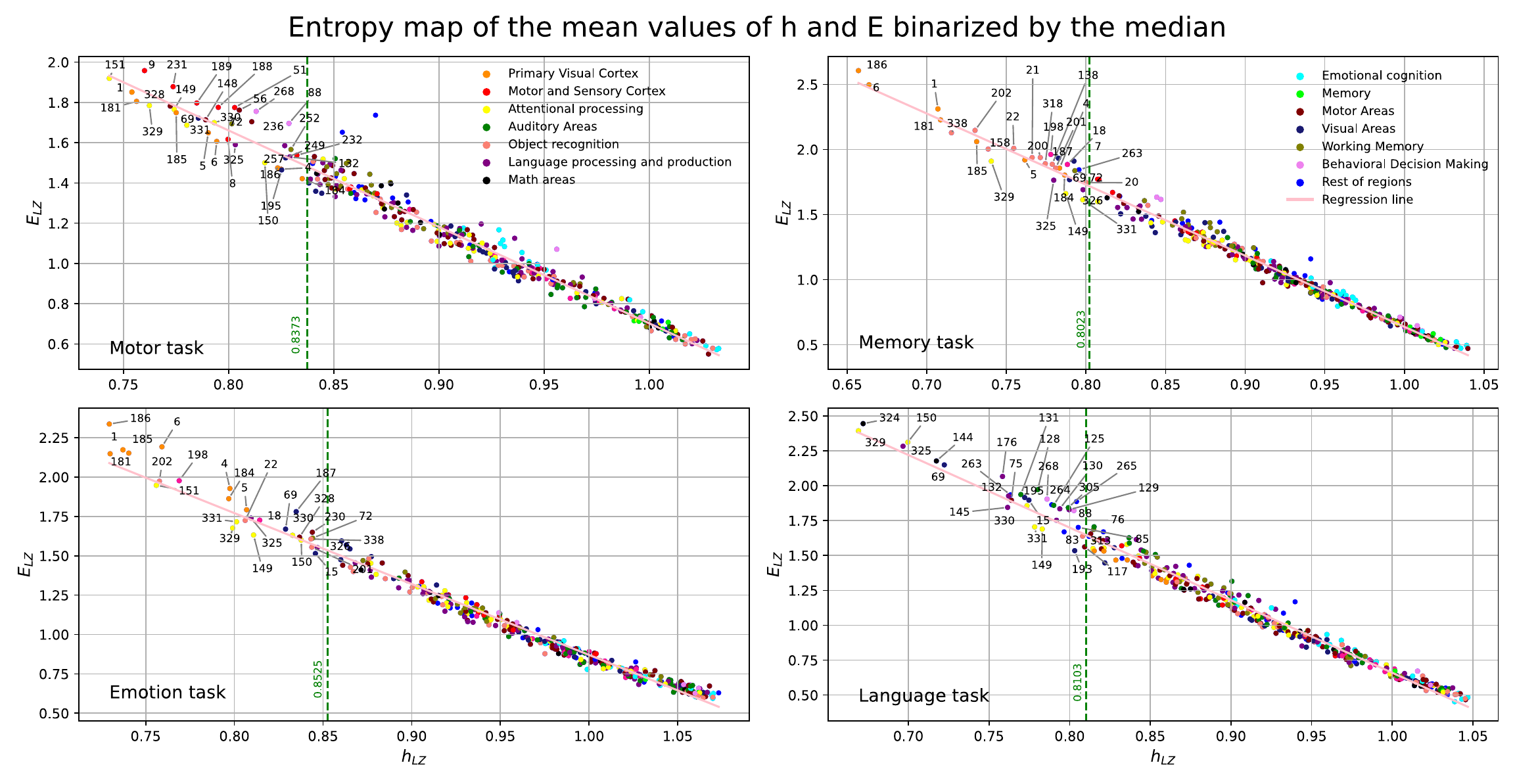}
    \caption{\textbf{Complexity entropy map under median binarization.} LZ effective measure complexity ($E_{LZ}$) versus LZ entropy density ($h_{LZ}$) for all 360 ROIs in the four tasks (Motor, Memory, Emotion, Language), computed exactly as in Figure 3 of the main text but binarizing each region's BOLD signal at its median instead of its mean. $h_{LZ}$ and $E_{LZ}$ were estimated per subject and averaged across subjects. The pink line is the least squares fit; the green dashed line is the activation threshold obtained with the previous residual criterion ($h_{LZ} = 0.8373$ motor, $0.8023$ memory, $0.8525$ emotion, $0.8103$ language). Point colors denote functional categories, and labels mark the active regions. Compared with the mean binarized maps, the points shift only slightly: the linear $E_{LZ}$–$h_{LZ}$ trend, the active/non-active classification, and the sign of each active region's residual relative to the regression line are preserved across all tasks.}
    \label{fig:Median_bin_entropic_map}
\end{figure}

The resulting entropy maps (Figure \ref{fig:Median_bin_entropic_map}) reproduce the mean binarized maps of Figure 3 (main text): individual points move slightly, but their relative ordering, the set of regions classified as active, and the sign of each active region's residual with respect to the regression line are unchanged across all four tasks. In the motor task the Primary and Sensory Motor Cortex remain above the regression line (over-performing) and the main visual areas below it (under-performing); in the emotion task the visual and face processing regions remain above the line. Both patterns match the main text result. The over/under-performance reported in the manuscript is therefore not an artifact of the mean threshold choice. This is consistent with the spectral entropy comparison (Supplementary Figure \ref{fig:SE_vs_hLZ}), which reaches the same conclusion without any binarization.

To quantify this agreement, we computed, across all 360 regions, the Spearman correlation between the $E_{LZ}$ residual relative to the regression line obtained under the mean threshold and the residual obtained under each alternative discretization (Table \ref{tab:threshold_robustness}). In addition to the median, we included a three level (tertile) quantization, in which each region's signal is mapped to three symbols using its 33rd and 66th percentiles, so as to test sensitivity to the size of the alphabet and not only to the position of the threshold. Under median binarization the residuals are strongly correlated with the mean threshold residuals in every task ($\rho = 0.86$-$0.93$, all $p < 10^{-100}$), confirming quantitatively the visual agreement of Figure \ref{fig:Median_bin_entropic_map}. The three level quantization, which the Lempel-Ziv estimator resolves less reliably at these sequence lengths because of the larger alphabet (see the discussion of ordinal encoding in the main text), gives a weaker but still strong agreement ($\rho = 0.62$-$0.79$, all $p < 10^{-38}$, which remain significant after Bonferroni correction). The classification and the over/under-performance are therefore robust to both the threshold choice and the alphabet size.

\begin{table}[ht!]
\centering
\caption{Robustness of the $E_{LZ}$ residual to the discretization choice. Spearman correlation $\rho$ between the per-region $E_{LZ}$ residual with respect to the regression line obtained under mean-threshold binarization and under two alternative discretizations: median-threshold binarization and three-level (tertile) quantization. The correlation is computed over all 360 regions; $p$ is the two-sided $p$-value. A high $\rho$ indicates that the over/under-performance structure of the complexity-entropy map is preserved.}
\label{tab:threshold_robustness}
\begin{tabular}{lcccc}
\toprule
 & \multicolumn{2}{c}{Median} & \multicolumn{2}{c}{Tertile} \\
\cmidrule(lr){2-3} \cmidrule(lr){4-5}
Task & $\rho$ & $p$ & $\rho$ & $p$ \\
\midrule
Motor    & 0.93 & $1.5\times10^{-154}$ & 0.74 & $8.2\times10^{-64}$ \\
Memory   & 0.93 & $7.3\times10^{-156}$ & 0.79 & $2.4\times10^{-77}$ \\
Emotion  & 0.86 & $1.4\times10^{-108}$ & 0.62 & $3.9\times10^{-39}$ \\
Language & 0.93 & $1.5\times10^{-154}$ & 0.76 & $1.5\times10^{-68}$ \\
\bottomrule
\end{tabular}
\end{table}

To check that the over/under-performance assignment of the individual active regions, and not only the global ordering, is robust to the discretization, we counted for each active set the number of regions whose $E_{LZ}$ residual keeps its sign under the median and tertile encodings, taking the mean threshold sign as reference (Table \ref{tab:sign_preservation}). The under-performing sets are fully preserved: the visual regions in the motor task and the attentional regions in the emotion task retain their sign in all regions under both encodings. The over-performing sets are largely preserved, with the emotion visual and face-processing set keeping its sign in 7 of 8 regions under the median and 6 of 8 under the tertile, and the motor and sensory cortex in 6 of 9 under both. The discrepancies are therefore confined to the over-performing assignment, the less robust of the two directions, while the global residual structure remains strongly correlated across discretizations (Table \ref{tab:threshold_robustness}).

\begin{table}[ht!]
\centering
\caption{Preservation of the over/under-performance under alternative discretizations. For each task and active set, the number of regions whose $E_{LZ}$ residual keeps its sign, and therefore its over- or under-performing assignment relative to the regression line, when the mean-threshold binarization is replaced by median binarization or by three level (tertile) quantization. The reference is the sign obtained under the mean threshold; $n$ is the number of regions in the set. The under-performing sets are fully preserved; the over-performing sets are largely preserved.}
\label{tab:sign_preservation}
\begin{tabular}{llccc}
\toprule
Task & Region set & Role & Median & Tertile \\
\midrule
Motor   & Primary and Sensory Motor Cortex & over ($+$)  & 6/9 & 6/9 \\
Motor   & Visual areas                     & under ($-$) & 8/8 & 8/8 \\
Emotion & Visual and face-processing       & over ($+$)  & 7/8 & 6/8 \\
Emotion & Attentional                      & under ($-$) & 8/8 & 8/8 \\
\bottomrule
\end{tabular}
\end{table}

\newpage

\subsection*{Supplementary information 4: Activation of visual/face processing versus attentional regions in the emotion task}

We tested whether the visual and face processing regions are more strongly engaged than the attentional regions in the emotion task, that is, whether they have lower $h_{LZ}$ and higher $E_{LZ}$. The visual/face set comprises the primary and early visual areas V1-V4 (ROIs 1/181, 4/184, 5/185, 6/186), the Fusiform Face Complex FFC (18/198), and the Inferior Temporal area PIT (22/202); the attentional set comprises PGs (151/331), PFm (149/329), PF (148/328) and PGi (150/330). The $h_{LZ}$ and $E_{LZ}$ values are the mean binarized estimates used in the main text.

The test accounts for a property of the data. The absolute level of the entropic map shifts from subject to subject, so a region's $h_{LZ}$ carries a subject specific offset in addition to its regional value; across the first ten subjects this offset spans $\overline{h}_{LZ} \approx 0.67$-$0.88$, larger than the visual/face - attentional gap itself. Because the offset is shared by all regions within a subject, it cancels in the within-subject difference between the two sets. We therefore reduced each subject to the difference between the mean of the visual/face set and the mean of the attentional set, separately for $h_{LZ}$ and $E_{LZ}$, and tested these $N = 153$ paired values across subjects. This is equivalent to a subject random intercept model in which the offset is absorbed by the random intercept and set membership is the fixed effect; the subject is the unit of replication.

\begin{table}[ht!]
\centering
\caption{Visual/face versus attentional regions in the emotion task, paired across the 153 subjects. For each measure the table reports the within-subject mean difference (visual/face $-$ attentional), its 95\% bootstrap confidence interval (5000 resamples), the paired $t$-test and Wilcoxon signed rank $p$-values, the standardized effect size (Cohen's $d_z$), and the number of subjects in which the difference is in the expected direction.}
\label{tab:emotion_vf_at}
\begin{tabular}{lcccccc}
\toprule
Measure & Mean diff. & 95\% CI & $p_{t}$ & $p_{\mathrm{W}}$ & $d_z$ & Expected dir. \\
\midrule
$h_{LZ}$ & $-0.049$ & $[-0.061,\,-0.037]$ & $1.5\times10^{-13}$ & $1.9\times10^{-12}$ & $-0.66$ & 114/153 \\
$E_{LZ}$ & $+0.333$ & $[+0.260,\,+0.405]$ & $6.6\times10^{-16}$ & $1.4\times10^{-14}$ & $+0.73$ & 122/153 \\
\bottomrule
\end{tabular}
\end{table}

Across the 153 subjects the visual/face regions have significantly lower $h_{LZ}$ and higher $E_{LZ}$ than the attentional regions (Table~\ref{tab:emotion_vf_at}); both effects are of medium size, are highly significant by both a parametric and a non-parametric test, and survive Bonferroni correction for the two comparisons. The Wilcoxon $p$-values ($1.9\times10^{-12}$ for $h_{LZ}$, $1.4\times10^{-14}$ for $E_{LZ}$) confirm the result without a normality assumption.

\begin{figure}[ht!]
    \centering
    \includegraphics[width=\textwidth]{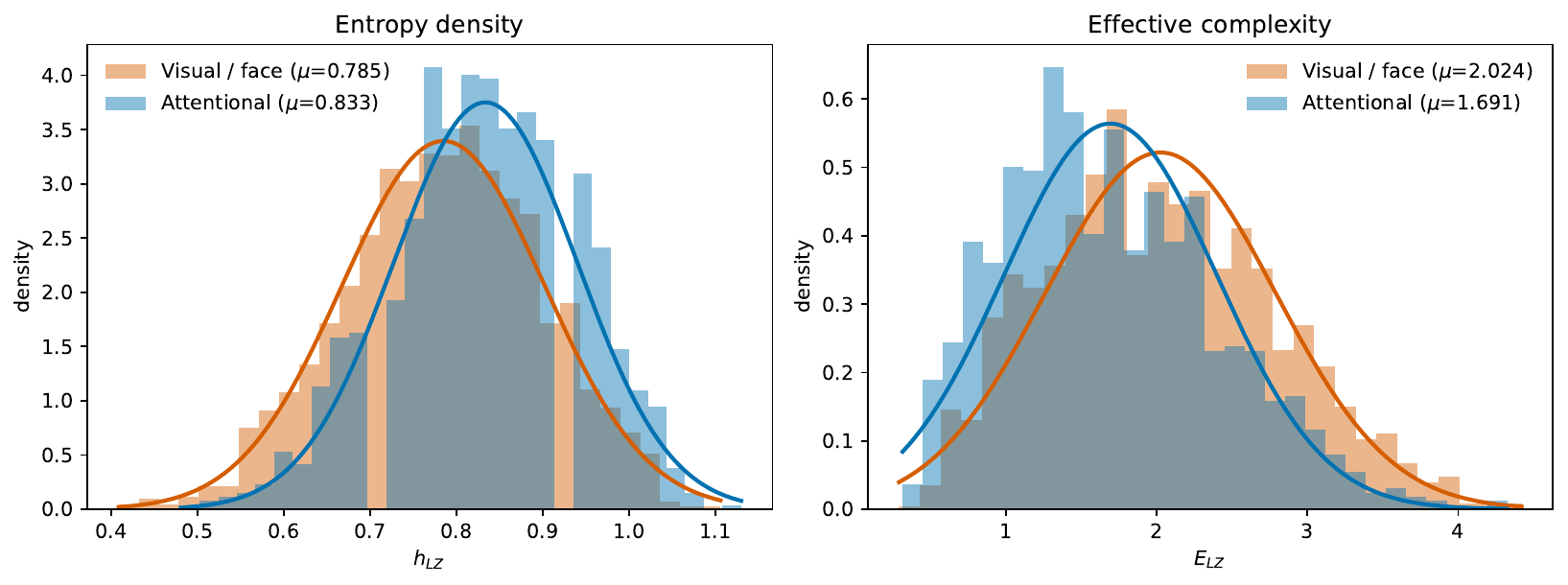}
    \caption{\textbf{Pooled distributions of $h_{LZ}$ and $E_{LZ}$ for the visual/face and attentional sets in the emotion task.} Each histogram pools all region$\times$subject values (visual/face $n = 12\times153$, attentional $n = 8\times153$); the solid curves are Gaussian fits with the sample mean and standard deviation. The visual/face set is shifted toward lower $h_{LZ}$ and higher $E_{LZ}$. The substantial overlap reflects the per-subject map offset, which adds between subject variance to the pooled distributions; this offset is removed in the within subject paired test (Table~\ref{tab:emotion_vf_at}), which is the basis for the inference.}
    \label{fig:emotion_vf_at_pooled}
\end{figure}

\begin{figure}[ht!]
    \centering
    \includegraphics[width=\textwidth]{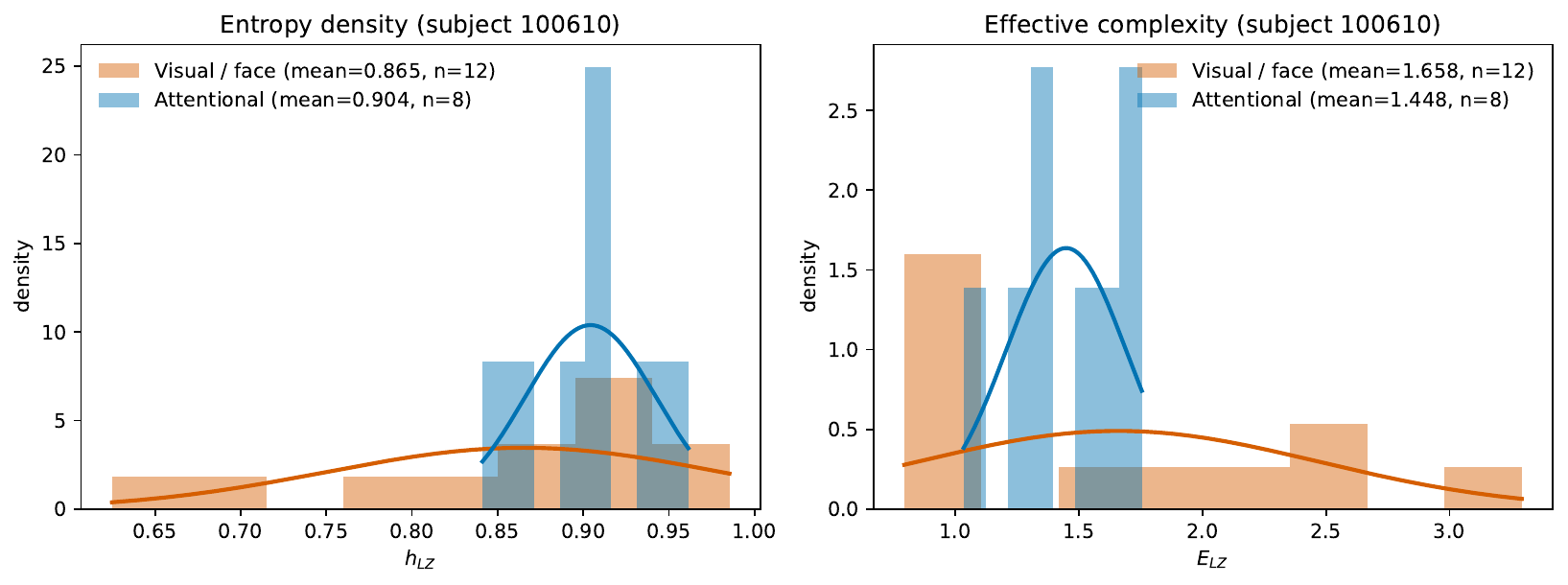}
    \caption{\textbf{Single-subject distributions (subject 100610, the first one, no selection).} $h_{LZ}$ (left) and $E_{LZ}$ (right) for the 12 visual/face and 8 attentional regions of one subject, with Gaussian overlays. The two sets differ in the expected direction but the separation is not resolvable within a single subject: the sets contain few, non-independent (bilaterally homologous) regions, so a per-subject test has negligible power (here Welch $p = 0.29$ for $h_{LZ}$ and $0.41$ for $E_{LZ}$), and the sign of the gap even reverses in some subjects. This is why the effect is established across subjects rather than within one.}
    \label{fig:emotion_vf_at_single}
\end{figure}

The pooled distributions (Figure~\ref{fig:emotion_vf_at_pooled}) show the shift in mean but overlap because the per-subject offset inflates their variance; the pooled effect sizes ($d = -0.33$ for $h_{LZ}$, $d = 0.39$ for $E_{LZ}$) therefore underestimate the within-subject effect. At the single-subject level (Figure~\ref{fig:emotion_vf_at_single}) the difference is in the expected direction but not statistically resolvable, which is expected from the small number of non-independent regions per set and motivates the across-subject paired design. The stronger engagement of visual and face-processing regions relative to attentional regions in the emotion task is thus a reliable group-level property, not an impression from the complexity-entropy map.

\newpage

\subsection*{Supplementary information 5: Symmetry across different tasks}

To complement the main analysis shown in Figure 5 (Main text), we include here the remaining tasks: Language, Motor, and Emotion, to illustrate the consistency of the observed interhemispheric symmetry across all stimulus conditions. As in the Memory task, these additional cases confirm that task-driven brain activity elicits a strong bilateral pattern, with homotopic regions in both hemispheres displaying similar entropy density levels. This supplementary figure further supports the contrast with the resting state, where symmetry breaks down and widespread asymmetry emerges.

\begin{figure*}[ht!]
        \centering
            \includegraphics[width=0.6\textwidth]{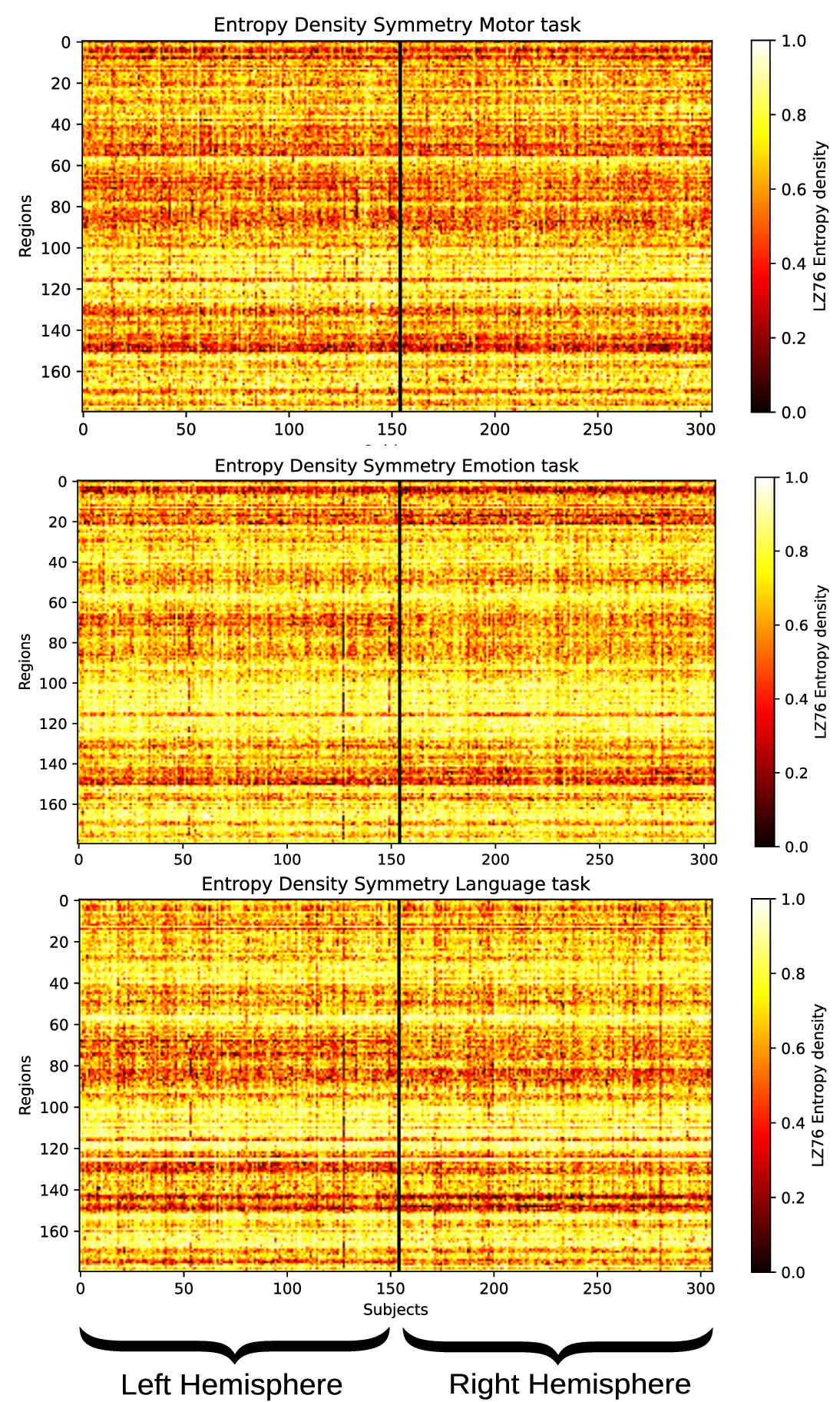}

            \caption{ \textbf{Symmetry across different tasks.} The figure shows the entropy density levels of each region across all subjects (N = 153) during the Motor, Emotion, and Language tasks. The left side displays regions in the left hemisphere (from 0 to 180), while the right side corresponds to the homologous regions in the right hemisphere (from 181 to 360). During each task performance, a clear symmetry is observed between hemispheres, with homologous regions exhibiting similar activation levels.}
            \label{fig:symmetry}
\end{figure*}

\newpage

\subsection*{Supplementary information 6: Network Enrichment and Significance}

\begin{figure*}[ht!]
        \centering
            \includegraphics[width=\textwidth]{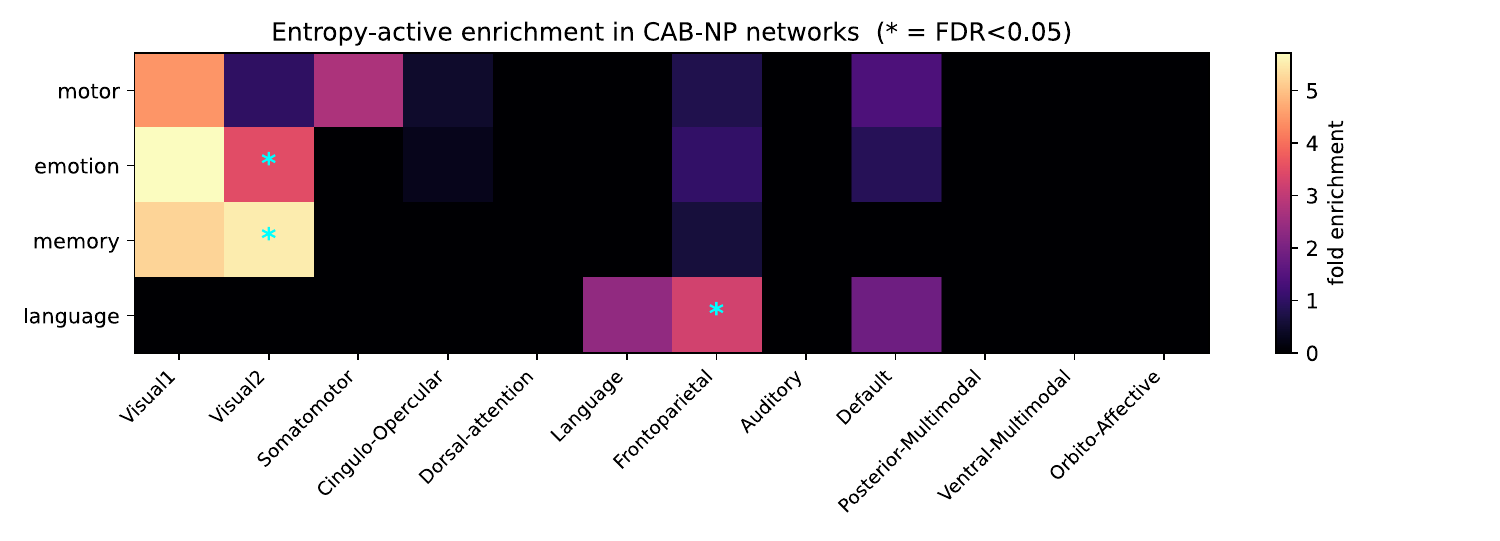}

            \caption{ \textbf{ Fold enrichment of the entropy-active set across all twelve Cole–Anticevic networks for the four tasks} heat map; asterisks mark FDR-significant enrichment}
            \label{fig:heatmap_fold_enrichment}
\end{figure*}

Network enrichment of the entropy-active regions in the Cole–Anticevic Brain-wide Network Partition is detailed explained in the following table \ref{tab:network_analysis} anf Figure \ref{fig:heatmap_fold_enrichment}. For each task and each of the twelve Cole–Anticevic networks, the table reports: network, the network name; overlap $(k)$, the number of entropy-active regions assigned to that network; net\_size $(K)$, the total number of cortical regions in the network; n\_active (n), the total number of active regions for the task; fold, the fold enrichment $(k/n)/(K/N)$ with $N = 360$, where 1 is the chance level; p, the raw hypergeometric p-value; q\_value, the Benjamini-Hochberg adjusted p-value (corrected across the twelve networks within each task); and fdr\_sig, whether the network is significant at $q \leq 0.05$. One network per task survives correction: Visual2 for working memory (fold $5.5, q = 7\times10^{-13})$ and emotion (fold $3.5, q = 4\times10^{-4}$), and Frontoparietal for language (fold $3.2, q = 6\times10{-3}$). For motor, the strongest enrichments are in the somatomotor (fold $2.7, p = 4\times10^{-3}$) and Visual1 (fold 4.4) networks; these do not survive correction across the twelve networks because the motor active set is split between the movement and visual-cue components, consistent with the GLM analysis.

\begin{table}[htbp]
\setlength{\tabcolsep}{1pt}
\centering
\caption{Task Network Enrichment and Significance}
\label{tab:network_analysis}
\begin{adjustbox}{width=\textwidth}
\begin{tabular}{l l c c c c S S c S}
\toprule
\textbf{task} & \textbf{network} & \textbf{net\_id} & \textbf{overlap} & \textbf{net\_size} & \textbf{n\_active} & \textbf{fold} & \textbf{p} & \textbf{fdr\_sig} & \textbf{q\_value} \\
\midrule
motor & Visual1 & 1 & 2 & 6 & 27 & 4.444444444444445 & 0.06741229460274278 & False & 0.4044737676164567 \\
motor & Visual2 & 2 & 4 & 54 & 27 & 0.9876543209876543 & 0.600182575185947 & False & 1.0 \\
motor & Somatomotor & 3 & 8 & 39 & 27 & 2.7350427350427347 & 0.00439832214145599 & False & 0.05277986569747188 \\
motor & Cingulo-Opercular & 4 & 2 & 56 & 27 & 0.47619047619047616 & 0.9445661013694068 & False & 1.0 \\
motor & Dorsal-attention & 5 & 0 & 23 & 27 & 0.0 & 1.0 & False & 1.0 \\
motor & Language & 6 & 0 & 23 & 27 & 0.0 & 1.0 & False & 1.0 \\
motor & Frontoparietal & 7 & 3 & 50 & 27 & 0.7999999999999999 & 0.7552564669840927 & False & 1.0 \\
motor & Auditory & 8 & 0 & 15 & 27 & 0.0 & 1.0 & False & 1.0 \\
motor & Default & 9 & 8 & 77 & 27 & 1.3852813852813852 & 0.19654696957906517 & False & 0.7861878783162607 \\
motor & Posterior-Multimodal & 10 & 0 & 7 & 27 & 0.0 & 1.0 & False & 1.0 \\
motor & Ventral-Multimodal & 11 & 0 & 4 & 27 & 0.0 & 1.0 & False & 1.0 \\
motor & Orbito-Affective & 12 & 0 & 6 & 27 & 0.0 & 1.0 & False & 1.0 \\
emotion & Visual1 & 1 & 2 & 6 & 21 & 5.714285714285714 & 0.042229023663835644 & False & 0.2533741419830139 \\
emotion & Visual2 & 2 & 11 & 54 & 21 & 3.4920634920634925 & 3.56939378663979e-05 & True & 0.00042832725439677474 \\
emotion & Somatomotor & 3 & 0 & 39 & 21 & 0.0 & 1.0 & False & 1.0 \\
emotion & Cingulo-Opercular & 4 & 1 & 56 & 21 & 0.3061224489795918 & 0.9743377255939446 & False & 1.0 \\
emotion & Dorsal-attention & 5 & 0 & 23 & 21 & 0.0 & 1.0 & False & 1.0 \\
emotion & Language & 6 & 0 & 23 & 21 & 0.0 & 1.0 & False & 1.0 \\
emotion & Frontoparietal & 7 & 3 & 50 & 21 & 1.0285714285714285 & 0.5795513236000547 & False & 1.0 \\
emotion & Auditory & 8 & 0 & 15 & 21 & 0.0 & 1.0 & False & 1.0 \\
emotion & Default & 9 & 4 & 77 & 21 & 0.8905380333951762 & 0.6939998561939772 & False & 1.0 \\
emotion & Posterior-Multimodal & 10 & 0 & 7 & 21 & 0.0 & 1.0 & False & 1.0 \\
emotion & Ventral-Multimodal & 11 & 0 & 4 & 21 & 0.0 & 1.0 & False & 1.0 \\
emotion & Orbito-Affective & 12 & 0 & 6 & 21 & 0.0 & 1.0 & False & 1.0 \\
memory & Visual1 & 1 & 2 & 6 & 23 & 5.217391304347826 & 0.05010412783689733 & False & 0.30062476702138397 \\
memory & Visual2 & 2 & 19 & 54 & 23 & 5.507246376811595 & 5.721483489524216e-14 & True & 6.865780187429059e-13 \\
memory & Somatomotor & 3 & 0 & 39 & 23 & 0.0 & 1.0 & False & 1.0 \\
memory & Cingulo-Opercular & 4 & 0 & 56 & 23 & 0.0 & 1.0 & False & 1.0 \\
memory & Dorsal-attention & 5 & 0 & 23 & 23 & 0.0 & 1.0 & False & 1.0 \\
memory & Language & 6 & 0 & 23 & 23 & 0.0 & 1.0 & False & 1.0 \\
memory & Frontoparietal & 7 & 2 & 50 & 23 & 0.6260869565217391 & 0.8577132218762584 & False & 1.0 \\
memory & Auditory & 8 & 0 & 15 & 23 & 0.0 & 1.0 & False & 1.0 \\
memory & Default & 9 & 0 & 77 & 23 & 0.0 & 1.0 & False & 1.0 \\
memory & Posterior-Multimodal & 10 & 0 & 7 & 23 & 0.0 & 1.0 & False & 1.0 \\
memory & Ventral-Multimodal & 11 & 0 & 4 & 23 & 0.0 & 1.0 & False & 1.0 \\
memory & Orbito-Affective & 12 & 0 & 6 & 23 & 0.0 & 1.0 & False & 1.0 \\
language & Visual1 & 1 & 0 & 6 & 20 & 0.0 & 1.0 & False & 1.0 \\
language & Visual2 & 2 & 0 & 54 & 20 & 0.0 & 1.0 & False & 1.0 \\
language & Somatomotor & 3 & 0 & 39 & 20 & 0.0 & 1.0 & False & 1.0 \\
language & Cingulo-Opercular & 4 & 0 & 56 & 20 & 0.0 & 1.0 & False & 1.0 \\
language & Dorsal-attention & 5 & 0 & 23 & 20 & 0.0 & 1.0 & False & 1.0 \\
language & Language & 6 & 3 & 23 & 20 & 2.347826086956522 & 0.12694496058191643 & False & 0.5077798423276657 \\
language & Frontoparietal & 7 & 9 & 50 & 20 & 3.2399999999999998 & 0.0004860067217807251 & True & 0.005832080661368701 \\
language & Auditory & 8 & 0 & 15 & 20 & 0.0 & 1.0 & False & 1.0 \\
language & Default & 9 & 8 & 77 & 20 & 1.8701298701298703 & 0.041510016173569766 & False & 0.2490600970414186 \\
language & Posterior-Multimodal & 10 & 0 & 7 & 20 & 0.0 & 1.0 & False & 1.0 \\
language & Ventral-Multimodal & 11 & 0 & 4 & 20 & 0.0 & 1.0 & False & 1.0 \\
language & Orbito-Affective & 12 & 0 & 6 & 20 & 0.0 & 1.0 & False & 1.0 \\
\bottomrule

\end{tabular}
\end{adjustbox}
\end{table}

\newpage

\subsection*{Supplementary information 7: Validation with GLM}
GLM validation (supporting Results, Section [validation]). Comparison of the entropic activation $(-h_{LZ})$ with the HCP group-average task GLM maps (Cohen's d, 997 subjects), within each task's functional network.

In Table~\ref{tab:correlation_results}, the bootstrap confidence intervals largely corroborate the significance tests. For motor, emotion, and working memory, the intervals exclude zero and remain moderately narrow, indicating stable positive correlations. In contrast, the language auditory result, although significant, exhibits a much wider interval due to the small number of regions $(n=15)$, suggesting that the existence of the effect is more certain than its precise magnitude. The full language network interval spans zero, providing no evidence for a reliable association.

\begin{table}[ht!]
\setlength{\tabcolsep}{1pt}
\centering
\begin{adjustbox}{width=\textwidth}
\begin{tabular}{l l l r S S S S r }
\hline
task & scope & network & n & rho & ci\_lo & ci\_hi & p & fdr\_sig \\
\hline
motor & all\_360 & - & 360 & 0.171272412081369 & 0.0678585919640934 & 0.268503646846023 & 0.00110432900314226 & False \\
motor & primary\_network & Somatomotor+Visual & 99 & 0.47334570191713 & 0.317889642256503 & 0.600205901225087 & 7.49817231783949E-07 & True \\
emotion & all\_360 & - & 360 & 0.252862033400463 & 0.143123852532962 & 0.351653172006299 & 1.17241513733083E-06 & False \\
emotion & primary\_network & Visual & 60 & 0.586162823006391 & 0.368052356468513 & 0.739717730266695 & 8.62138839376251E-07 & True \\
memory & all\_360 & - & 360 & 0.453558798035993 & 0.363364519467683 & 0.537164282692706 & 1.14612854903913E-19 & False \\
memory & primary\_network & Visual & 60 & 0.670408446790775 & 0.506300845376722 & 0.788673445662438 & 4.6420609819341E-09 & True \\
language & all\_360 & - & 360 & 0.210799350998037 & 0.112997553828948 & 0.305573161558217 & 5.55131545418488E-05 & False \\
language & primary\_network & Language+Frontoparietal+Auditory & 88 & -0.0605825789862985 & -0.273967404739129 & 0.159865695367372 & 0.574999018582938 & False \\
language & primary\_network & Auditory & 15 & 0.660714285714286 & 0.0471778656126483 & 0.912568306010929 & 0.00733056983089171 & True \\
\hline
\end{tabular}
\end{adjustbox}
\caption{\textbf{GLM–entropy correlation per task.} Spearman correlation between the entropic activation ($-h_{LZ}$) and the GLM effect size, computed over all 360 regions (all\_360) and within each task's functional network (primary\_network). Columns: task; scope; network; n, number of regions; rho, Spearman correlation; ci\_lo, ci\_hi, $95\%$ bootstrap confidence interval (5000 resamples); p, two-sided p-value; fdr\_sig, significance at $q \leq 0.05$ after Benjamini-Hochberg correction across the four primary network tests. The whole cortex correlations are weak $(\rho \in [0.17, 0.45])$ because the around 300 regions outside the task network carry no task signal and dilute the estimate. Within the task network the correlation is strong for motor ($\rho = 0.47$), emotion ($\rho = 0.59$) and working memory ($\rho = 0.67$). For language, the auditory network alone gives $\rho = 0.66$ while the full network gives $\rho = -0.06$; both are listed.}
\label{tab:correlation_results}
\end{table}

\begin{figure*}[ht!]
        \centering
            \includegraphics[width=\textwidth]{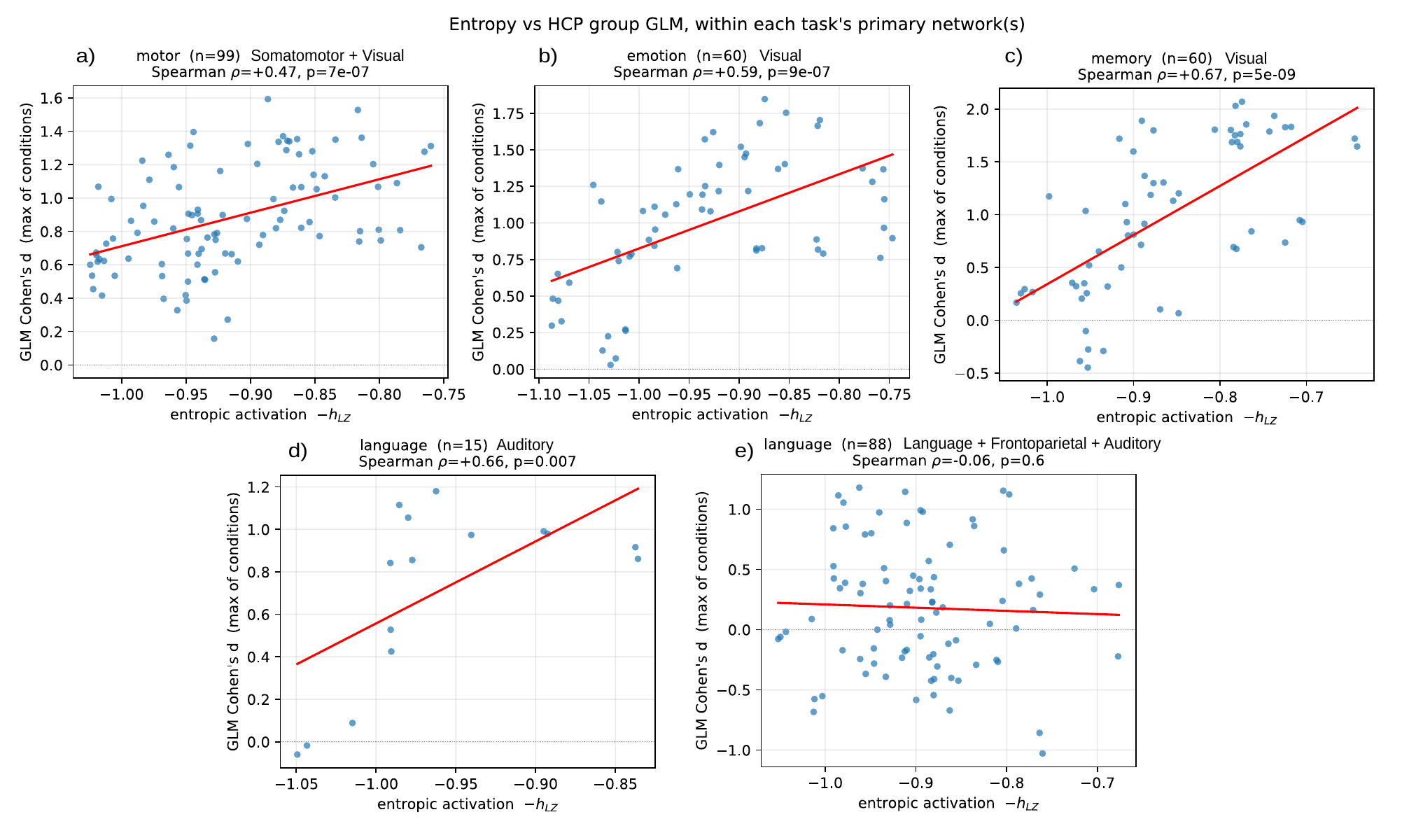}

            \caption{ \textbf{ Entropic activation versus GLM effect size per task}. Each point is one cortical region. The horizontal axis is the entropic activation ($-h_{LZ}$), the vertical axis the GLM effect size (Cohen's d, per region maximum of the task versus baseline condition contrasts). Points are restricted to the task's functional network (Cole-Anticevic partition): a) somatomotor and visual for motor, visual for b) emotion and c) working memory, auditory for d) language. The red line is the linear fit and the dotted line marks $d = 0$. Spearman $\rho$ and $p$ are given in each panel. For the language task two panels are shown: d) the auditory network alone ($\rho = 0.66, p = 7\times10^{-3}$) and e) the full Language + Frontoparietal + Auditory network ($\rho = -0.06, p = 0.57$); the difference is explained in Supplementary Table~\ref{tab:language_network_effects}.}
            \label{fig:scatter_Spearman}
\end{figure*}

\begin{table}[p]
\centering
\small
\begin{tabular}{llSS[table-format=3.2, round-mode=places, round-precision=2]r}
\hline
{task} & {network} & {mean\_cohens\_d} & {$\%$\_positive} & {n} \\
\hline
motor & Visual1 & 0.903110315402349 & 100 & 6 \\
motor & Visual2 & 0.798232463774858 & 100 & 54 \\
motor & Somatomotor & 0.994251020443745 & 100 & 39 \\
motor & Cingulo-Opercular & 1.03566716025983 & 100 & 56 \\
motor & Dorsal-attention & 1.13010871799096 & 100 & 23 \\
motor & Language & 0.619172920999932 & 100 & 23 \\
motor & Frontoparietal & 0.528006283761934 & 92 & 50 \\
motor & Auditory & 0.49781247874101 & 100 & 15 \\
motor & Default & 0.0425215363079174 & 61.038961038961 & 77 \\
motor & Posterior-Multimodal & 0.926058309418814 & 100 & 7 \\
motor & Ventral-Multimodal & 0.291479125618935 & 100 & 4 \\
motor & Orbito-Affective & 0.122004650998861 & 83.3333333333333 & 6 \\
\hline
emotion & Visual1 & 0.456976552183429 & 100 & 6 \\
emotion & Visual2 & 1.0587314263814 & 100 & 54 \\
emotion & Somatomotor & -0.0295077871101407 & 38.4615384615385 & 39 \\
emotion & Cingulo-Opercular & 0.089809531204602 & 53.5714285714286 & 56 \\
emotion & Dorsal-attention & 0.682358003828836 & 100 & 23 \\
emotion & Language & 0.176722020713066 & 69.5652173913043 & 23 \\
emotion & Frontoparietal & 0.0838510133326054 & 66 & 50 \\
emotion & Auditory & 0.110250693031897 & 73.3333333333333 & 15 \\
emotion & Default & -0.145387656865762 & 28.5714285714286 & 77 \\
emotion & Posterior-Multimodal & 0.452016848538603 & 100 & 7 \\
emotion & Ventral-Multimodal & 0.319709151983261 & 100 & 4 \\
emotion & Orbito-Affective & 0.0701697481175264 & 66.6666666666667 & 6 \\
\hline
memory & Visual1 & 0.401144180446863 & 83.3333333333333 & 6 \\
memory & Visual2 & 1.04377600316096 & 92.5925925925926 & 54 \\
memory & Somatomotor & -0.0886280749327852 & 30.7692307692308 & 39 \\
memory & Cingulo-Opercular & 0.14181140105107 & 53.5714285714286 & 56 \\
memory & Dorsal-attention & 0.98291626183883 & 100 & 23 \\
memory & Language & 0.155809600695806 & 56.5217391304348 & 23 \\
memory & Frontoparietal & 0.604367597363889 & 90 & 50 \\
memory & Auditory & -0.347868010401726 & 6.66666666666667 & 15 \\
memory & Default & -0.163127149107515 & 32.4675324675325 & 77 \\
memory & Posterior-Multimodal & 0.249888181154217 & 85.7142857142857 & 7 \\
memory & Ventral-Multimodal & 0.412064522504807 & 100 & 4 \\
memory & Orbito-Affective & -0.00828583398833871 & 66.6666666666667 & 6 \\
\hline
language & Visual1 & -0.249000093278786 & 16.6666666666667 & 6 \\
language & Visual2 & -0.310683188272243 & 0 & 54 \\
language & Somatomotor & -0.28727873089986 & 5.12820512820513 & 39 \\
language & Cingulo-Opercular & -0.166685022580038 & 26.7857142857143 & 56 \\
language & Dorsal-attention & 0.129311700473013 & 56.5217391304348 & 23 \\
language & Language & 0.403153774508214 & 82.6086956521739 & 23 \\
language & Frontoparietal & -0.0817888213135302 & 42 & 50 \\
language & Auditory & 0.71556205401818 & 86.6666666666667 & 15 \\
language & Default & -0.2576505029711 & 16.8831168831169 & 77 \\
language & Posterior-Multimodal & -0.25058130999761 & 0 & 7 \\
language & Ventral-Multimodal & -0.00903915427625179 & 25 & 4 \\
language & Orbito-Affective & -0.222613346452514 & 0 & 6 \\
\hline
\end{tabular}
\caption{\textbf{Mean GLM effect size per network for all four tasks}. Mean GLM Cohen's d and percentage of positive-d regions in each of the twelve Cole-Anticevic networks for all four tasks. Columns: network; mean\_cohens\_d, mean effect size across the network's regions; $\%$\_positive, percentage of regions with $d > 0$; n, number of regions in the network. In the Language task, the GLM activates the Auditory (mean $d = +0.72$, $87\%$ positive) and Language (mean $d = +0.40$, $83\%$ positive) networks, but the Frontoparietal network is task-negative (mean $d = -0.08$, $42\%$ positive). The frontoparietal regions are engaged by the arithmetic blocks of the language run, but relative to the run's baseline they fall below it, so the GLM represents them as deactivated. This is why the full network correlation is null while the auditory network correlation is strong, and why these regions are instead recovered by the network-enrichment and Neurosynth validations. For the other three tasks, each task's primary network shows positive mean $d$ (motor, emotion, working memory)}
\label{tab:language_network_effects}
\end{table}

\newpage

\subsection*{Supplementary information 8: Inter-subject variability and individual-versus-group reproducibility}

The complexity-entropy maps, distance matrices and dendrograms reported in the main text are group-level summaries: $h_{LZ}$ and $E_{LZ}$ were computed for each of the 360 regions in each of the $N=153$ subjects after binarization, and the plotted values are the across-subject means; the distance matrices and dendrograms are the means of the per-subject LZ distance matrices. Here we quantify how variable these quantities are across subjects and whether the activation pattern is present at the individual level or emerges only after averaging.

Table~\ref{tab:variability} reports, per task: $N$, the number of subjects; $\mathrm{CV}(h_{LZ})$ and $\mathrm{CV}(E_{LZ})$, the median across the 360 regions of the across-subject coefficient of variation (standard deviation divided by the mean) of each measure, in percent; $\mathrm{SEM}$ is the median across-region standard error of the mean, as a percentage of the mean ($\mathrm{SEM}=\mathrm{CV}/\sqrt{N}$) and $\rho_{h}$ and $\rho_{E}$, the Spearman correlation between each subject's 360-region profile and the group-mean profile, averaged over subjects and given as mean\,$\pm$\,SD. The latter was computed leave-one-subject-out, so that the reference profile for each subject is the mean of the remaining 152 and does not contain the subject itself. 

\begin{table}[ht!]
\centering
\caption{Inter-subject variability and individual-versus-group reproducibility of the entropic measures. $\mathrm{CV}$ is the median across-region coefficient of variation across subjects; $\mathrm{SEM}$ is the median across-region standard error of the mean, as a percentage of the mean ($\mathrm{SEM}=\mathrm{CV}/\sqrt{N}$); $\rho_h$, $\rho_E$ are the mean$\,\pm\,$SD Spearman correlation of each subject's profile with the leave-one-out group mean. Median SEM percentages are additionally reported for both measures.}
\label{tab:variability}
\begin{tabular}{lccccccc}
\toprule
Task & $N$ & $\mathrm{CV}(h_{LZ})$ [\%] & $\mathrm{CV}(E_{LZ})$ [\%] & $\rho_{h}$ & $\rho_{E}$ & \makecell{Median\\SEM$_h$ [\%]} & \makecell{Median\\SEM$_E$ [\%]}\\
\midrule
Motor    & 153 & 8.8 & 51.3 & $0.70 \pm 0.09$ & $0.54 \pm 0.09$ & 0.714 & 4.151 \\
Memory   & 153 & 8.0 & 53.2 & $0.76 \pm 0.07$ & $0.63 \pm 0.07$ & 0.650 & 4.299 \\
Emotion  & 153 & 7.6 & 50.0 & $0.72 \pm 0.08$ & $0.53 \pm 0.09$ & 0.618 & 4.039 \\
Language & 153 & 9.1 & 53.4 & $0.75 \pm 0.07$ & $0.63 \pm 0.08$ & 0.739 & 4.313 \\
\bottomrule
\end{tabular}
\end{table}

\begin{figure}[ht!]
\centering
\includegraphics[width=0.9\textwidth]{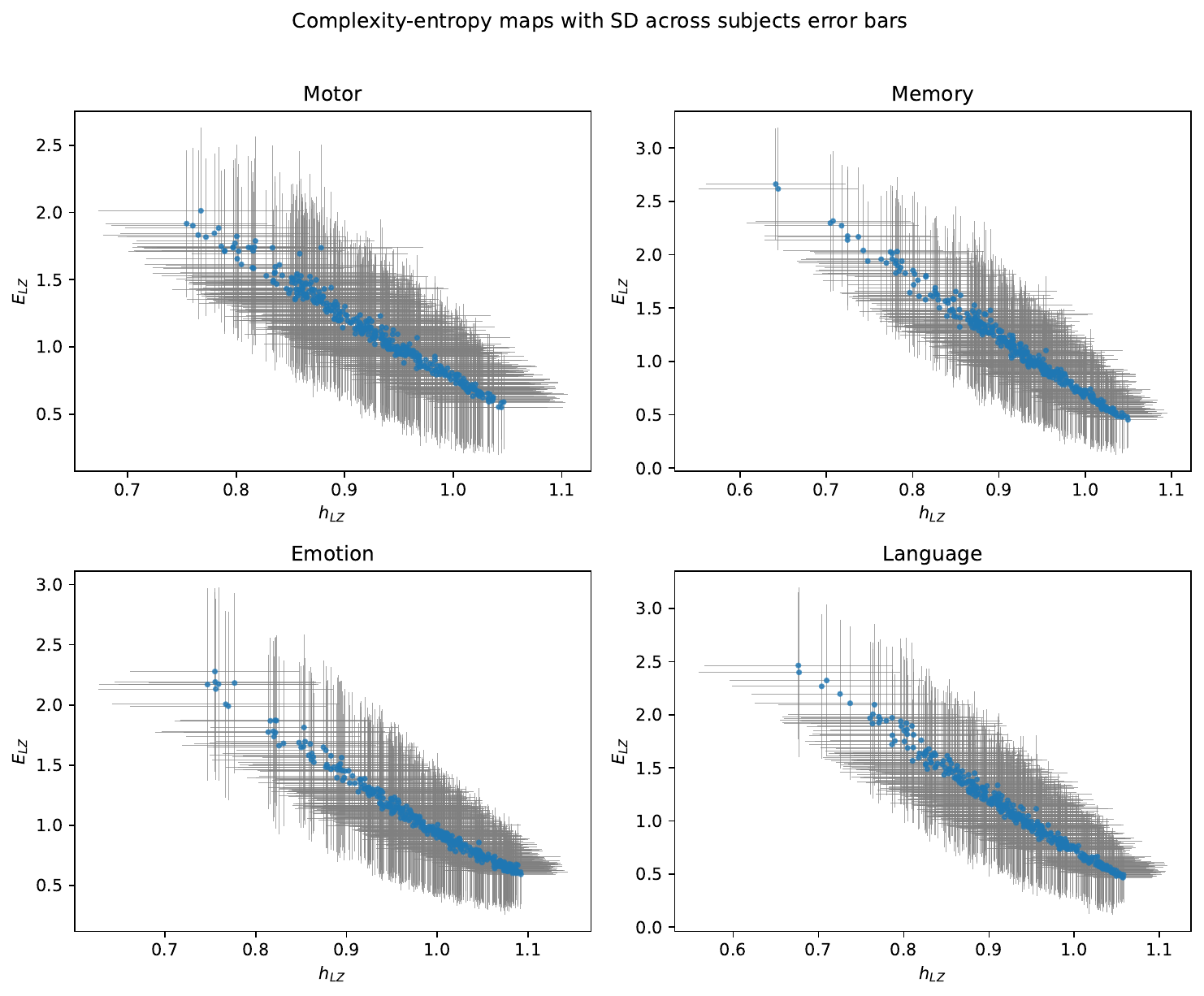}
\caption{Complexity-entropy maps with across-subject error bars. Each point is the group mean $(h_{LZ}, E_{LZ})$ of one region; horizontal and vertical bars are the across-subject standard deviation. The active regions, at the upper-left of each panel, are separated from the bulk mainly along $h_{LZ}$, where the inter-subject spread is small relative to the separation.}
\label{fig:errorbars}
\end{figure}

The two coordinates behave differently. The entropy density $h_{LZ}$ is stable across subjects, with a median coefficient of variation between 7.6\% and 9.1\% across tasks. Because the quantity plotted and thresholded is the group mean, its uncertainty is smaller by a factor $\sqrt{N}$: the standard error of the mean $h_{LZ}$ is on the order of 0.6-0.7\% of its value, so the position of each region along the axis used to separate active from non-active regions is tightly determined (Figure~\ref{fig:errorbars}, SEM version). The effective measure complexity $E_{LZ}$, estimated by the block-shuffling procedure, is intrinsically noisier per subject, with a median coefficient of variation of about 50-53\%; the corresponding standard error of the group mean is approximately 4\%, so the mean $E_{LZ}$ remains well estimated even though single-subject values are not. This is the reason the over/under-performance relative to the regression line, which is read on the $E_{LZ}$ axis, is supported by an across-subject residual test rather than by any single map.

The activation pattern is reproducible at the individual level. Each subject's $h_{LZ}$ profile correlates with the group-mean profile with $\rho_{h}$ between 0.70 and 0.76, so an individual map resembles the group map closely; the lower $\rho_{E}$ (0.53-0.63) again reflects the larger per-subject noise of $E_{LZ}$.

\newpage

\subsection*{Supplementary information 9: Distribution across subjects of the most and least active region}

\begin{figure}[ht!]
\centering
\includegraphics[width=\textwidth]{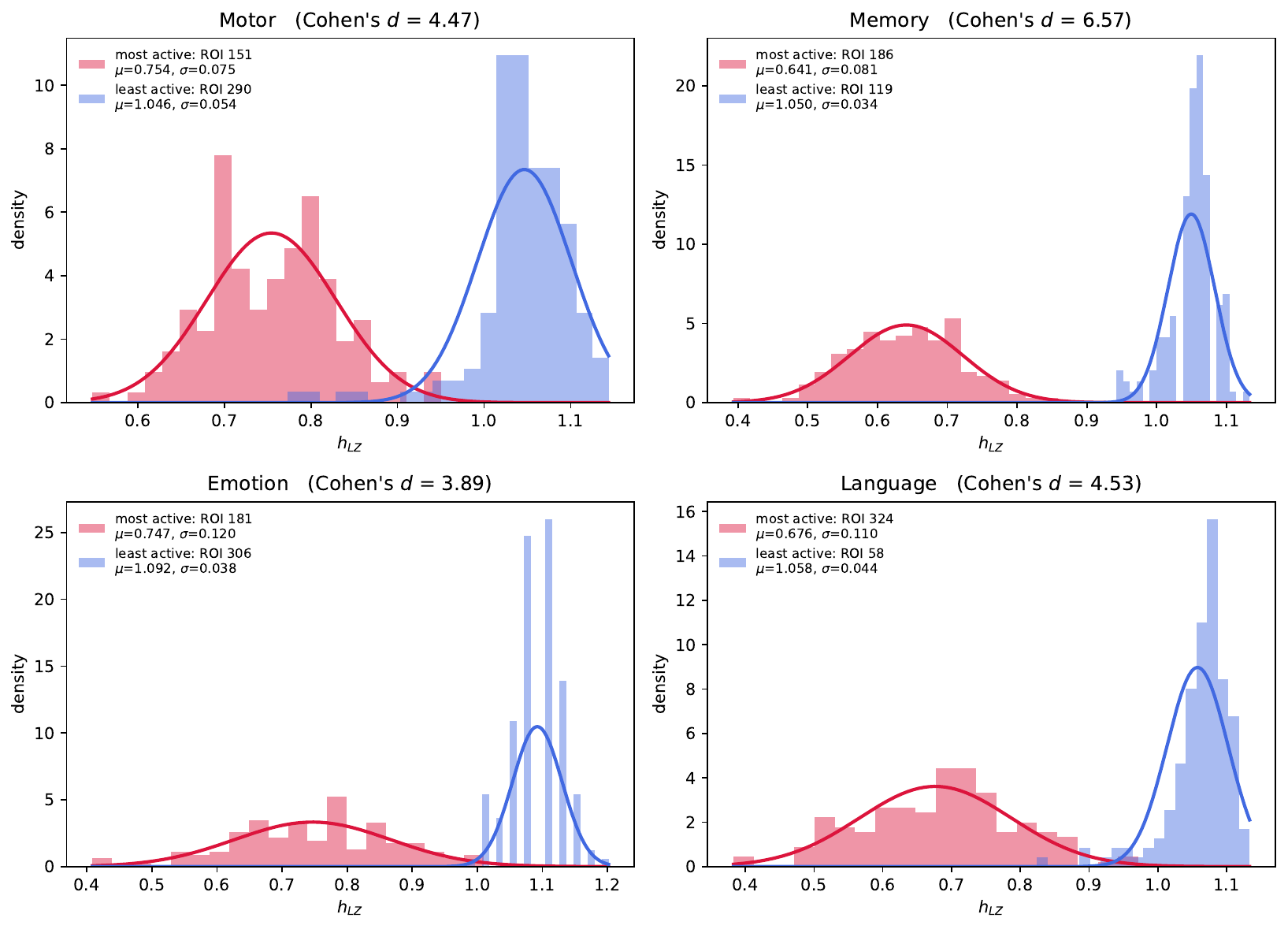}
\caption{\textbf{Across-subject distribution of $h_{LZ}$ for the most active (red) and least active (blue) region of each task, with Gaussian fits.} Histograms are normalized to unit area; curves are normal distributions with the sample mean and standard deviation of each region. Each panel reports the two means and standard deviations and the between-distribution Cohen's $d$.}
\label{fig:dist_active_inactive}
\end{figure}

For each task we identified the region with the lowest mean $h_{LZ}$ across subjects (the most active) and the region with the highest mean $h_{LZ}$ (the least active), and plotted the distribution of their $h_{LZ}$ values over the 153 subjects, with a normal distribution fitted to each by its sample mean and standard deviation (Figure~\ref{fig:dist_active_inactive}).

The two distributions are well separated in every task. The mean $h_{LZ}$ of the most active region lies between 0.64 and 0.75, against 1.05-1.09 for the least active region, and the standardized separation is large throughout (Cohen's $d = 4.47$ motor, $6.57$ memory, $3.89$ emotion, $4.53$ language; all far above the conventional large-effect value of 0.8). The fitted normals overlap only in their tails, so the strongly active and low active assignment of these regions holds at the level of individual subjects and is not produced by averaging.

\newpage

\subsection*{Supplementary information 10: Dendrograms.}

\begin{figure}[ht!]
    \centering
    \includegraphics[width=\textwidth]{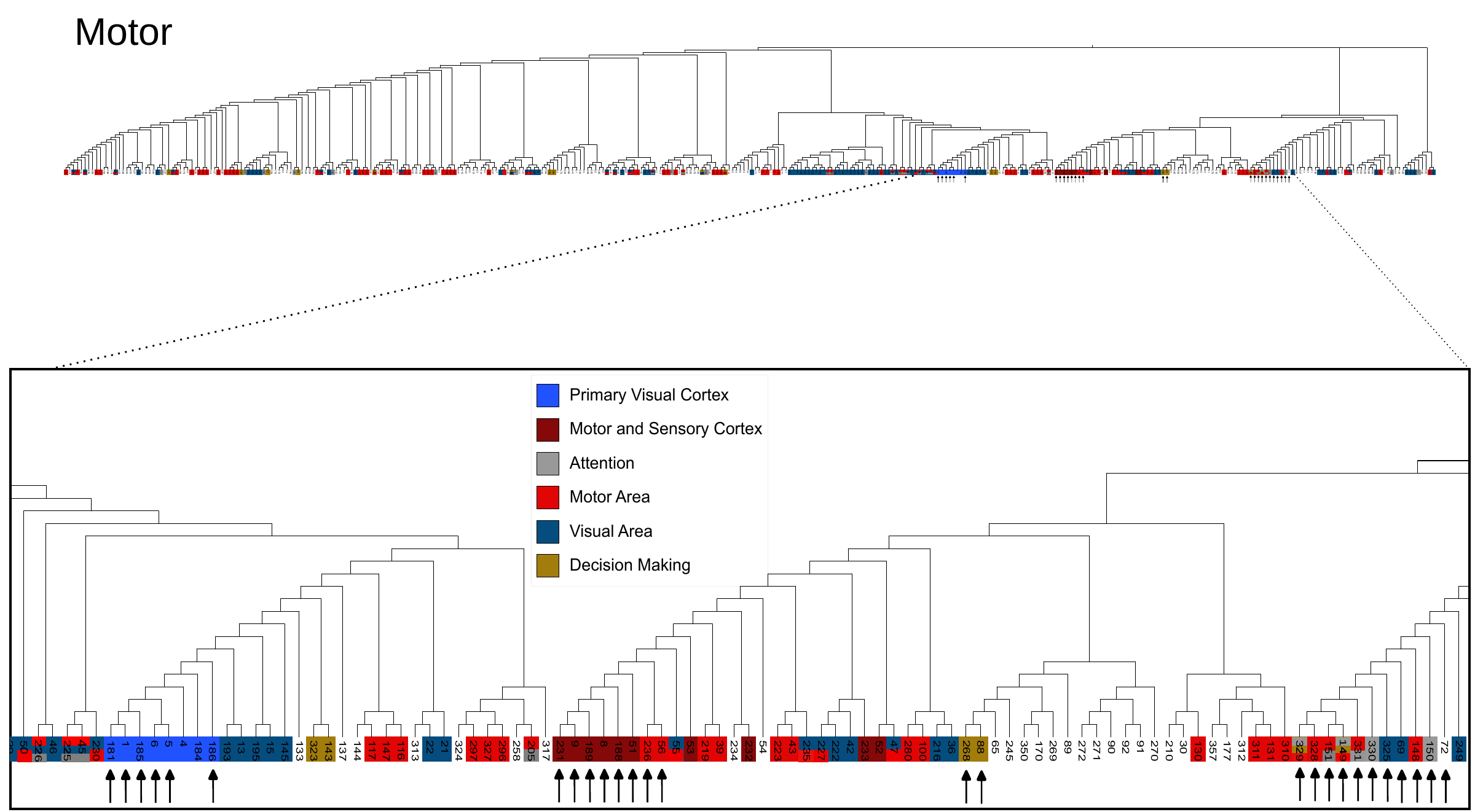}
    \caption{\textbf{Motor dendrogram.} A dendrogram was constructed from the distance matrix to illustrate hierarchical connections. Arrows point to active regions, showing clusters of visual, motor, and attentional processing regions, with strong associations among active regions. }
    \label{fig:motor_dendro}
\end{figure}

\begin{figure}[ht!]
    \centering
    \includegraphics[width=\textwidth]{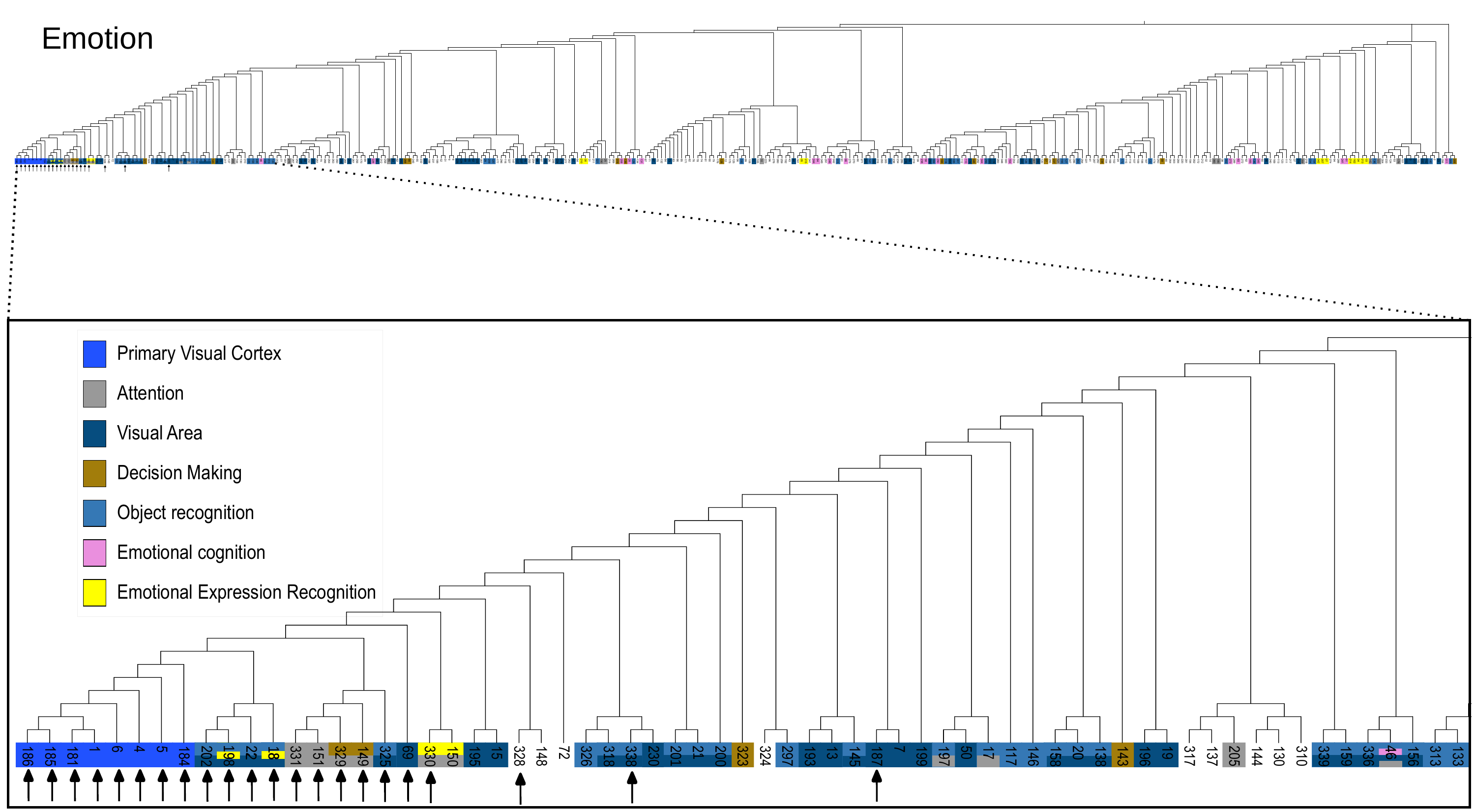}
    \caption{\textbf{Emotion dendrogram.} A dendrogram was constructed from the distance matrix to illustrate hierarchical connections. Arrows point to active regions, showing clusters of visual, motor, and attentional processing regions, with strong associations among active regions. }
    \label{fig:emotion_dendro}
\end{figure}

\begin{figure}[ht!]
    \centering
    \includegraphics[width=\textwidth]{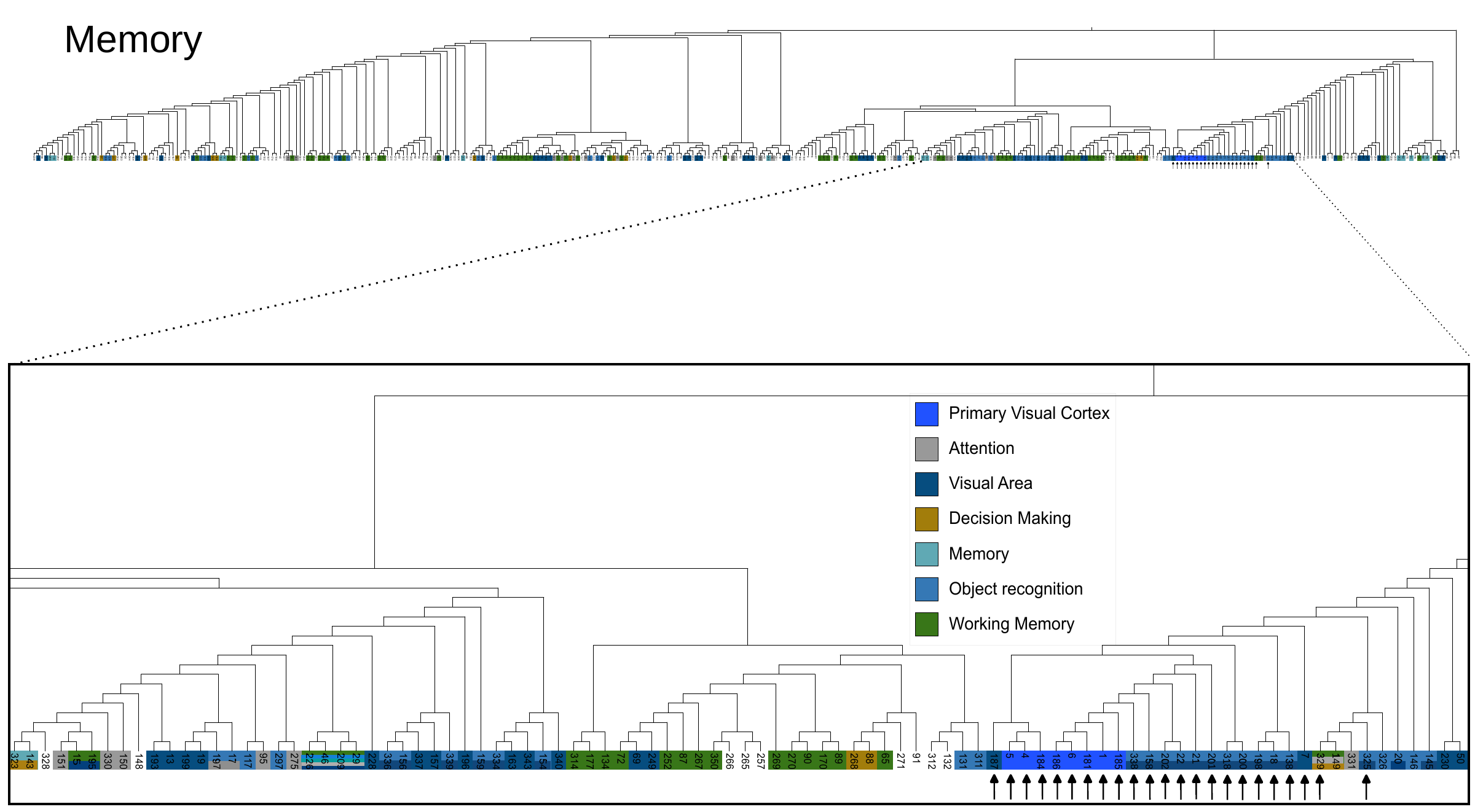}
    \caption{\textbf{Memory dendrogram.} A dendrogram was constructed from the distance matrix to illustrate hierarchical connections. Arrows point to active regions, showing clusters of visual, motor, and attentional processing regions, with strong associations among active regions. }
    \label{fig:memory_dendro}
\end{figure}

\begin{figure}[ht!]
    \centering
    \includegraphics[width=\textwidth]{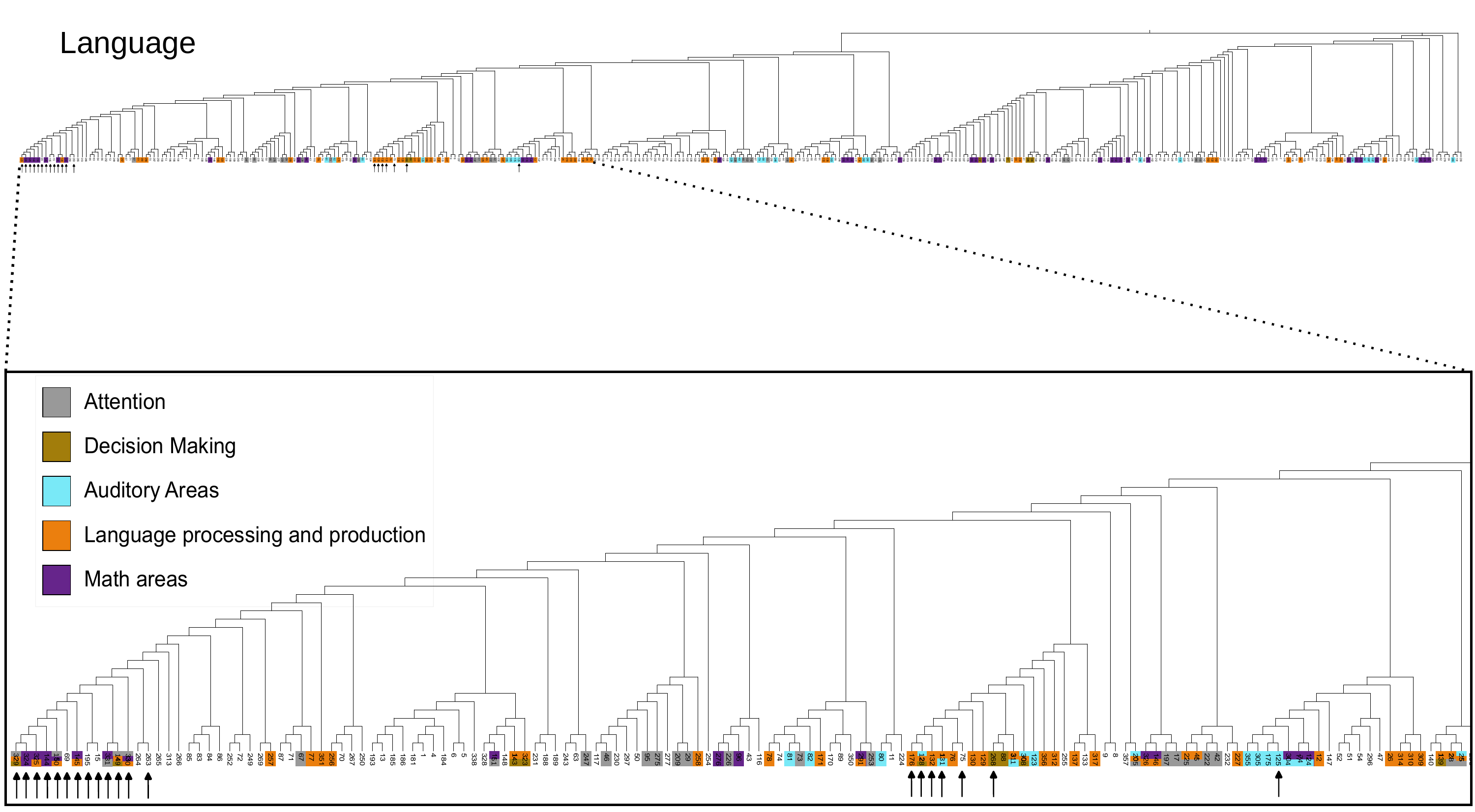}
    \caption{\textbf{Language dendrogram.} A dendrogram was constructed from the distance matrix to illustrate hierarchical connections. Arrows point to active regions, showing clusters of visual, motor, and attentional processing regions, with strong associations among active regions. }
    \label{fig:language_dendro}
\end{figure}

\newpage

\subsection*{Supplementary information 11: Validation of the dendrograms}

We quantified two properties of the LZ-distance dendrograms: the faithfulness of each tree to its distance matrix, and the reproducibility of its cluster structure across subjects.

The cophenetic correlation is the Pearson correlation between the leaf to leaf path distances in the Neighbor-Joining tree and the original LZ-distances; a high value indicates that the tree represents the distance matrix without substantial distortion. The subject level bootstrap resamples the 153 subjects with replacement (500 resamples), rebuilds the across subject mean distance matrix and the Neighbor-Joining tree for each resample, and records, for every pair of regions that share a cluster in the full sample partition, the fraction of resamples in which they remain co-clustered (co-clustering support). Each resample draws 153 subjects with replacement, so it has the same size as the original sample. The full-sample partition and each resampled tree were cut into $k = 12$ flat clusters, matching the twelve Cole–Anticevic networks used in the enrichment analysis (Supplementary Section 6); the support reported is stable under moderate changes of $ k$.
\begin{table}[ht!]
\centering
\caption{Validation of the LZ-distance dendrograms. Cophenetic correlation between the tree path distances and the LZ-distances, and mean within cluster co-clustering support from a subject level bootstrap (153 subjects, 500 resamples).}
\label{tab:dendro_validation}
\begin{tabular}{lcc}
\toprule
Task & Cophenetic corr. & Bootstrap support \\
\midrule
Motor    & 0.83 & 0.82 \\
Emotion  & 0.89 & 0.68 \\
Memory   & 0.87 & 0.72 \\
Language & 0.89 & 0.73 \\
\bottomrule
\end{tabular}
\end{table}

The cophenetic correlations (0.83-0.89) show that the dendrograms faithfully represent the distance matrices. The bootstrap support (0.68-0.82) shows that the cluster structure is reproducible under subject resampling. The correspondence of the clusters to canonical functional systems is established separately by the network enrichment analysis (Supplementary Section 6), in which the active regions of each task are significantly enriched in the expected Cole-Anticevic network.

\bibliography{My_Library2}

\end{document}